\documentclass[
 aip,
 apr,
 amsmath,amssymb,
 reprint
]{revtex4-1}

\usepackage{graphicx}
\usepackage{dcolumn}
\usepackage{bm}

\usepackage[utf8]{inputenc}
\usepackage[T1]{fontenc}
\usepackage{mathptmx}

\graphicspath{{./Figures/}}
\usepackage{multirow}
\usepackage{amssymb}

\usepackage[english]{babel}

\usepackage{siunitx}
\usepackage{textcomp}
\usepackage{hyperref}
\hypersetup{
    colorlinks,
    citecolor=black,
    filecolor=black,
    linkcolor=black,
    urlcolor=black
}

\usepackage{natbib}

\begin{document}

\title{Compact gate-based read-out of multiplexed quantum devices with a cryogenic CMOS active inductor}

\author{L.~Le~Guevel}
\email{loick.leguevel@gmail.com}
\affiliation{Univ. Grenoble Alpes, CEA, LETI, F-38000 Grenoble, France}
\affiliation{Univ. Grenoble Alpes, CEA, Grenoble INP, IRIG, PHELIQS, F-38000 Grenoble, France}
\author{G.~Billiot}
\affiliation{Univ. Grenoble Alpes, CEA, LETI, F-38000 Grenoble, France}
\author{S.~De~Franceschi}
\affiliation{Univ. Grenoble Alpes, CEA, Grenoble INP, IRIG, PHELIQS, F-38000 Grenoble, France}
\author{A.~Morel}
\affiliation{Univ. Grenoble Alpes, CEA, LETI, F-38000 Grenoble, France}
\author{X. Jehl}
\author{A.G.M.~Jansen}
\affiliation{Univ. Grenoble Alpes, CEA, Grenoble INP, IRIG, PHELIQS, F-38000 Grenoble, France}
\author{G.~Pillonnet}
\affiliation{Univ. Grenoble Alpes, CEA, LETI, F-38000 Grenoble, France}

\date{\today}

\begin{abstract}
In the strive for scalable quantum processors, significant effort is being devoted to the development of cryogenic classical hardware for the control and readout of a growing number of qubits.
Here we report on a cryogenic circuit incorporating a CMOS-based active inductor enabling fast impedance measurements with a sensitivity of 10~aF and an input-referred noise of 3.7~aF/sqrt(Hz).
This type of circuit is especially conceived for the readout of semiconductor spin qubits.
As opposed to commonly used schemes based on dispersive rf reflectometry, which require mm-scale passive inductors, it allows for a markedly reduced footprint (50µm~×~60µm), facilitating its integration in a scalable quantum-classical architecture.
In addition, its active inductor results in a resonant circuit with tunable frequency and quality factor, enabling the optimization of readout sensitivity.
\end{abstract}

\maketitle

An ingenious use of the laws of quantum mechanics has led to a new computing paradigm, generally known as quantum computing, that promises exponential speed-up in the solution of certain types of problems\cite{shor_polynomial-time_1996, grover_quantum_1997, georgescu_quantum_2014}.
Using a prototypical quantum processor with 53 operational superconducting quantum bits (qubits), a ground-breaking experiment was recently  able to perform a first experiment towards quantum supremacy\cite{arute_quantum_2019}, triggering more extensive research on such a goal\cite{zhou_what_2020}.
Practical implementations of quantum computing, however, are expected to require much larger numbers of physical qubits\cite{gidney_how_2019}.

Solid-state implementations seem to offer the best scalability prospects. While superconducting qubits are currently the leading platform, semiconductor spin qubits are emerging as a serious contender owing to the possibility to leverage the integration capabilities of silicon technology\cite{maurand_cmos_2016}.
In both cases, the quantum processor can only function at very low temperature, typically below 0.1 K (only recently, it was shown that silicon qubits could be operated even above 1 K with limited loss of fidelity\cite{urdampilleta_gate-based_2019,yang_operation_2020}).

In a scale-up prospect toward increasingly large numbers of qubits, the introduction of classical cryogenic electronics positioned as close as possible to the qubits is widely considered as a necessity.
Various solutions have been proposed and partly demonstrated to a first proof-of-concept level.
These include low-temperature (de)multiplexers\cite{ward_integration_2013,paquelet_wuetz_multiplexed_2020}, analog-to-digital and digital-to-analog converters\cite{homulle_cryogenic_2016,bardin_design_2019}, low-noise amplifiers\cite{montazeri_ultra-low-power_2016,patra_cryo-cmos_2018}, RF oscillators\cite{guevel_110mk_2020,gong_193_2020}, transimpedance amplifiers\cite{tagliaferri_modular_2016,le_guevel_low-power_2020}, and digital processors\cite{van_dijk_scalable_2020,schriek_cryo-cmos_2020}.
In a DRAM-like strategy\cite{schaal_cmos_2019,ruffino_integrated_2021}, these cryogenic components can significantly reduce the number of electrical lines running through the host cryostat, thereby limiting the associated heat load and increasing interconnect reliability.
A CMOS-based cryogenic controller operating  at 3K was recently reported enabling high-fidelity operations on an electron-spin  two-qubit  system\cite{xue_cmos-based_2020}.
As  far as  qubit readout  is concerned, however, relatively little  has been  done.
Measuring the qubit state requires detecting small variations in the impedance of an LC tank circuit coupled to the qubit, which is commonly  done through  rf reflectometry. 
The inductive element of this tank circuit consists of a surface-mount inductor or a microfabricated superconducting  coil. Even for this second case, the corresponding  footprint  is relatively  large ($\sim\si{\milli\meter^2}$)  and  hence hardly  compatible with large-scale  qubit integration.

Here we propose an alternative readout technique involving a cryogenic CMOS-based active inductor with a compact design. 
Besides its reduced footprint favoring scalability,  this  CMOS inductor  offers  the  possibility to tune  the  characteristic  frequency  and  the  quality factor  of  the resonator,  which is instrumental in optimizing measurement  sensitivity.
As opposed to conventional reflectometry, our method consists in measuring the tank impedance at resonance.
The cryogenic read-out circuit is composed  of a current source exciting the LC tank, an amplifier to read the voltage response, and a multiplexed capacitor bank to select different devices under test (DUT).
We characterized the circuit sensitivity and tunability at 4.2 K demonstrating its capability to measure capacitances as low as 10 aF.
By applying our technique to a gate-coupled MOSFET co-integrated on the same chip we reveal typical signatures of quantized electronic states.

\begin{figure}[tb]
    \includegraphics{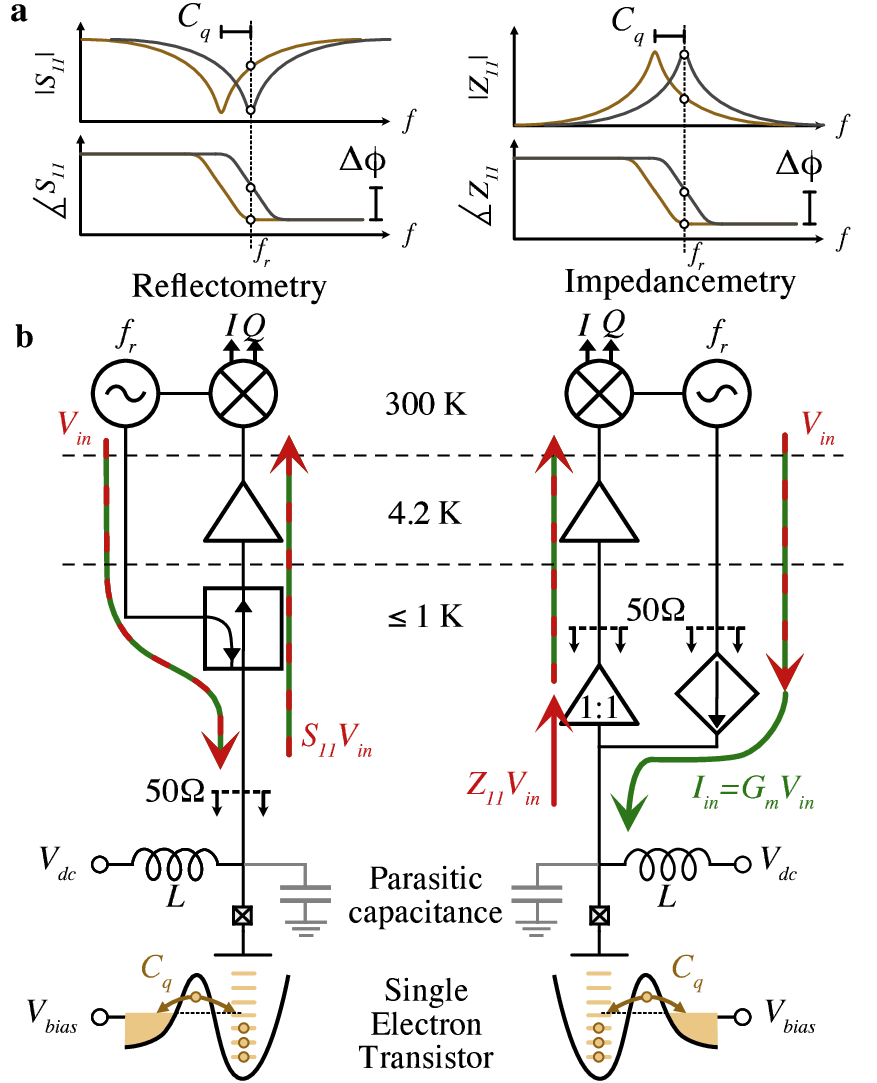}
    
    \caption{\label{fig:refimp}\textbf{Integration of measuring circuitry for scalable read-out of quantum capacitance. a}, Schematic signals of the complex scattering coefficient $S_{11}$ and impedance $Z_{11}$ of a resonant circuit in, respectively, reflectometry and impedancemetry. 
        \textbf{b}, Comparison between a typical reflectometry setup (left) and the proposed impedancemetry setup (right) for the measurement of the quantum capacitance $C_q$ of a single-electron transistor.
    Impedancemetry leverages cryogenic electronics to achieve higher integration of the measurement circuitry by getting rid of bulky directional couplers.
        Red (respectively green) arrows represent voltages (resp. currents).
        Red-green arrows emphasize the voltage-current interdependence due to signal propagation in \SI{50}{\ohm} lines.}

\end{figure}

\section{Impedancemetry}

Capacitive spectroscopy of gate-controlled quantum-dot devices allows the detection of electronic quantum states within the structure, including the firstly occupied electron states.
For enhanced detection sensitivity at high speed, the gate capacitance of the device under test (DUT), represented as a single-electron transistor in Figure~\ref{fig:refimp}b, is connected to an inductance to form an LC resonator.
In a gate-coupled read-out scheme of the quantum state, the response of the tank at cryogenic temperature excited near resonance frequency $f_r$ is conveyed back to room temperature for probing and analysis, usually with homodyne I-Q detection.
The phase of the tank response becomes an image of the change in DUT capacitance $\Delta C\ll C_p$ (see Figure~\ref{fig:refimp}a) through the relation $\Delta\phi=Q\Delta C/C_p$ around $f_r$ with $Q$ the resonator quality factor and $C_p$ the parallel capacitance.

Figure~\ref{fig:refimp}b shows a schematic comparison for resonance experiments between the commonly used method of reflectometry and the here employed method of impedancemetry.
Reflectometry uses voltages to excite and probe the resonator via the scattering\cite{crippa_gate-reflectometry_2019} or transmission parameters\cite{zheng_rapid_2019}, based on propagating waves.
To isolate the incoming and outgoing signals, directional couplers or circulators are used.

Impedancemetry uses currents to excite and probe the resonator via the impedance parameters without the need of bulky coupling elements.
The incoming signal $V_{in}$ at the resonant frequency $f_r$, generated at room temperature, is converted in a current $I_{in}=G_mV_{in}$ with a voltage-controlled current source of transimpedance $G_m$  at the base-temperature stage.
The input current $I_{in}$ creates a voltage $V_{out}=Z_rI_{in}$ through the tank impedance $Z_r$ that carries the information about the DUT capacitance.
$V_{out}$ is conveyed to a low-power unity-gain amplifier (follower) placed nearby the DUT to reduce parasitic capacitance $C_{par}$.
Main amplification is placed at a higher temperature (typically \SI{4.2}{\kelvin}) to benefit from higher cooling power.

Impedancemetry has the advantage over reflectometry that the \SI{50}{\ohm} impedance matching plays no role in the optimization of the resonant circuit depending on the inductor and the parasitic capacitors.
However, the cryogenic circuitry required by impedancemetry generates extra noise compared to reflectometry, which needs to be minimized.
The impedance of the resonator naturally filters out-of-resonance components (see Figure~\ref{fig:refimp}a) such as low-frequency flicker noise from electronics.
In the perspective of quantum computing involving a qubit matrix, $V_{in}$ could contain a comb of excitation frequencies to excite a set of frequency-selective resonators (see Supplementary Material~I with an estimation of the scaling for large qubit arrays).

In the case of impedancemetry, without the need of directional couplers, the footprint of the read-out circuitry is reduced.
Using modern CMOS technologies with sub-100nm nodes, the additional circuitry of current sources and followers easily fits on a chip with size comparable to the  hundreds of qubits chip (<\si{\milli\meter^2}) such that the total footprint is limited by the size of passive \si{\micro\henry}~inductance occupying a few \si{\milli\meter^2}.
The connection fan-out of a qubit matrix, originating from different-scale objects, increases the average interconnection length, thus lowers the detection sensitivity with important parasitic capacitance.
The applied high magnetic field ($\sim\SI{1}{\tesla}$) required to separate spin states via the Zeeman effect prevents an effective use of ferrite materials for reducing the inductance size. 
Active inductances consisting of transistors and capacitors can reach an inductance density as high as a few \si{\milli\henry/\milli\meter^2} which is 3 to 4 orders of magnitude higher than passive inductances (see Supplementary Material~I for the scaling analysis).
In the following, important issues of dissipation and noise of the realized active inductance will be treated in relation to sensitive capacitance detection.

\section{Active inductance}

The active inductance behavior is realized by transforming a capacitor $C_L$ into an inductance $L=C_L/G_{m,1}G_{m,2}$ via two transistor devices of transimpedance $G_{m,1}$ and $G_{m,2}$ forming a gyrator\cite{yuan_cmos_2008}. 
The non-ideal finite conductance and parasitic capacitance of the transistors set the resonant frequency $f_r$  and quality factor $Q$.
More advanced active inductance architectures incorporate a negative resistor in parallel to the inductance in order to improve $Q$ up to a few hundred with independent tuning of the inductance value $L$ and the quality factor $Q$\cite{yuan_cmos_2008, karsilayan_high-frequency_2000}. 

Fine calibration of the tunable inductance value using variable capacitors leads to a precise definition of the resonant frequency value, ideal for optimal frequency-multiplexing of large qubit matrices.
The tunability of the $Q$-factor enables different modes of read-out. 
High-$Q$ gives a precise measurement of quantum capacitance to calibrate qubit matrices.
Lower-$Q$ is more suitable for fast read-out during quantum computation.

\section{Integrated circuit design}

\begin{figure*}[tb]
    \includegraphics{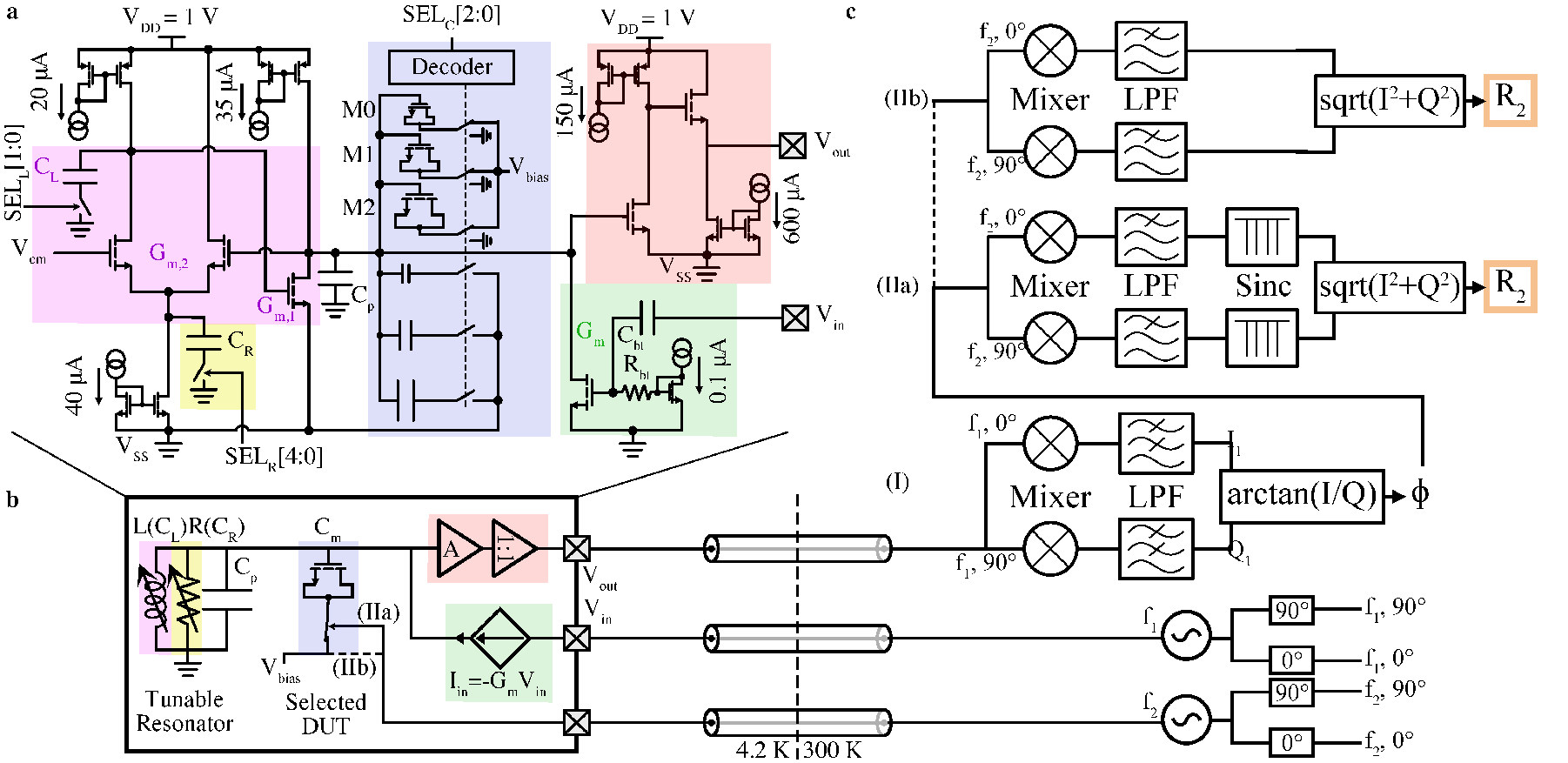}
    \caption{\label{fig:elecsetup}\textbf{Setup with on-chip electronics. a}, On-chip circuit implementation of the active inductance (pink), current excitation (green), test capacitor bank (blue), and amplification stage (red). For clarity, the bias MOSFETs operating in DC are drawn of smaller size than MOSFETs in the high-frequency signal chain.
        \textbf{b}, Simplified view of the on-chip resonant circuit placed at \SI{4.2}{\kelvin} with tunable resonator, DUT, current excitation, and voltage amplification, linked to room-temperature phase-sensitive electronics via meter-long cables.
    \textbf{c}, Room-temperature homodyne detection with single (I) and double (II) demodulation of the circuit output $V_{out}$ and generation of voltage excitation $V_{in}$ at modulation frequencies $f_1$ (150-\SI{200}{\mega\hertz}) and $f_2$ (\SI{1}{\kilo\hertz}).}
\end{figure*}

The impedancemetry experiment was integrated on a single chip with multiplexed quantum devices using the Fully-Depleted Silicon-On-Insulator (FD-SOI) 28nm CMOS technology.
The FD-SOI technology is very well suited for high-speed cryogenic applications\cite{nyssens_28-nm_2020} with lower variability than bulk technologies\cite{paz_integrated_2020}, less sensitivity to carrier freeze-out, and threshold-voltage tuning with the use of the back-gate\cite{paz_variability_2020}.
The integration of classical circuitry with small-enough transistors that exhibit quantum properties is a plus to validate efficiently new circuit architectures\cite{le_guevel_low-power_2020, guevel_110mk_2020,bonen_cryogenic_2019}

The realized integrated circuit contains the current source, the active inductance with addressable capacitor banks for tunability, the multiplexed DUTs, and the amplification stage (Figure~\ref{fig:elecsetup}a, b).
During design, we focused on bringing down the footprint and power consumption of the active inductance being the main original component of our circuit.
The complete amplification and current generation was added on-chip to facilitate testing the concept of impedancemetry at \SI{4.2}{\kelvin}.
In the absence of high-frequency models of FD-SOI transistors at cryogenic temperatures, we designed the integrated circuit with accurate room-temperature models supplied by the foundry\cite{poiroux_leti-utsoi21_2015}.
The evolution of transistor characteristics towards the lowest temperatures was extrapolated from the temperature variation in foundry models but also from acquired \SI{4.2}{\kelvin} data of single transistors\cite{paz_integrated_2020,paz_variability_2020}.

The active inductance follows a known NMOS-based Karsilayan-Schaumann architecture\cite{xiao_54-ghz_2007, barthelemy_nmos_2003}.
The gyrator is made of a single-ended negative transconductance $-G_{m,1}$ and a differential transconductance stage $G_{m,2}$.
The gyrator transforms a tunable capacitance $C_L$ into an inductance $L(C_L)=C_L/G_{m,1}G_{m,2}$.
An added metal-oxide-metal (MOM) capacitor $C_p$ of \SI{136}{\femto\farad} parallel to $L$ controls the resonant frequency $f_r=1/2\pi\sqrt{L(C_L)C_p}$.
No dependence in temperature is expected for MOM capacitors\cite{patra_characterization_2019}.
Adding $C_p$ makes the measuring circuit less sensitive to the DUT-capacitance with the increased tank capacitance but avoids the influence of unknown parasitic capacitances at cryogenic temperatures (e.g. substrate parasitics).
Hence, the resonant frequency $f_r$ is set by $C_p$, $C_L$, and $G_{m,i=1,2}$.
$C_L$ is implemented with one main metal-oxide-metal (MOM) capacitor of \SI{362}{\femto\farad} in parallel with two digitally-controlled binary-weighted MOM capacitors of 68 and \SI{136}{\femto\farad} (see Supplementary Material~II).
At room temperature, the emulated $L$ ranges from 5.3 to \SI{8.4}{\micro\henry} to reach $f_r$ from 128 to \SI{165}{\mega\hertz} (see Supplementary Material~III).
The estimated power consumption of the resonator is  $\SI{85}{\micro\watt}$ and corresponds to a footprint of \SI{8.5}{\micro\watt/qubit} assuming a  $10\times 10$ frequency-multiplexed array of qubits with frequency resolution of \SI{5}{\mega\hertz} in a \SI{1}{\giga\hertz} bandwidth.

Adding a capacitance $C_R$ at the foot of the differential transconductance stage allows to introduce a negative resistance in series with the active inductance, leading to higher Q-factor with an increased parallel effective resistance $R(C_R,C_L)$.
The $Q$-factor of the active inductance defined as $Q=R(C_L, C_R)\sqrt{L(C_L)/C_p}$ depends on $C_R$ and $C_L$.
By tuning $C_L$, then $C_R$, $L$ and $Q$ can be adjusted to any desired value apart from possible instabilities.
To tune the $Q$ factor, we choose to cover a wide range of $C_R$ values in steps of \SI{23}{\femto\farad} by selecting 4 binary-weighted MOM capacitors of 23, 46, 92, and \SI{184}{\femto\farad}.
From room-temperature simulations, these settings allow to cover a wide range of quality factors $Q$ from 7 to 300, including the unstable states with negative $Q$ (see Supplementary Material~III).

The voltage-controlled current source exciting the resonator is made of a current mirror combined to an RC bias tee.
The bias tee superimposes DC signals from the diode transistor to set the DC operating point of the current source and AC signals from the excitation input $V_{in}$ to generate the AC current $I_{in}$.
The RC filter of the bias tee consists of $R_{bt}$ (polysilicon resistor of \SI{10}{\mega\ohm}) and $C_{bt}$ (MOM capacitor of \SI{406}{\femto\farad}) to reach a characteristic frequency of \SI{39}{\kilo\hertz}.
As no large signals $V_{in}$ are required, the current source operates in subthreshold regime with a bias current of only \SI{0.1}{\micro\ampere} to minimize its conductance for a negligible impact on the resonator and obtain a desirable low transconductance for nA-current excitation. 
From foundry models, we get for the current generating transistor a transconductance $G_m$ of \SI{3.4}{\nano\ampere/\milli\volt} and bandwidth of \SI{3.5}{\giga\hertz} (see Supplementary Material~III).

To investigate the active inductance circuit with different DUTs, we added an addressable bank of 6 capacitors.
Three MOM capacitors of 2, 4, and \SI{8}{\femto\farad} have the purpose of calibrating the active inductance on known values.
Three additional MOSFETs (M0, M1, M2) of width \SI{80}{\nano\meter} and length 28, 60, and \SI{120}{\nano\meter} are used as test-bench for the investigation of quantum properties.
The source and drain voltage of the quantum MOSFETs are grounded when unselected and polarized at $V_{bias}$ when selected.
The differential transconductance stage of the active inductance copies the DC common-mode voltage $V_{cm}$ to the DUT gate potential, such that the DC gate voltage $V_{gs}=V_{cm}-V_{bias}$ can be varied via $V_{bias}$ (see Figure~\ref{fig:elecsetup}).

Once excited by $I_{in}$, the tank voltage is amplified, then sent through a unit-gain buffer for detection at room temperature via meter-long cable.
The amplifier is a common-source N-type single-stage and the 1:1 buffer is a common-drain N-type single stage (see Figure~\ref{fig:elecsetup}a).
Based on room-temperature simulations, the amplifier has a gain $A$ of \SI{15}{\deci\bel} and a  bandwidth of \SI{1.8}{\giga\hertz} for a power consumption of \SI{150}{\micro\watt}.
The buffer reaches a bandwidth of \SI{92}{\mega\hertz} for a cable capacitance of \SI{50}{\pico\farad} and a power consumption of \SI{2.4}{\milli\watt}.
The net amplification at \SI{165}{\mega\hertz} becomes \SI{8}{\deci\bel}.

Transistor noise translates into transconductance noise that generates perturbing variations into the parameters determining the active inductance. 
A varying $L$ modulates $f_r$ and generates phase noise in $V_{out}$.
The phase noise spectrum of $V_{out}$ around the carrier frequency $f_r$ extracted from room-temperature steady-state simulations (SST) exhibits a flicker component (see Supplementary Material~III) on time-scale  >\SI{10}{\milli\second}, induced by a noisy modulated $L$.
For a typical $Q$ of 81 with sufficiently fast measurements to avoid $1/f$ noise, we get a phase noise of \SI{0.002}{\degree/\sqrt{\hertz}} that gives an input-referred noise of \SI{3.2}{\atto\farad/\sqrt{\hertz}}.

Voltage excitation and homodyne detection are performed at room temperature with an all-digital lock-in amplifier (Figure~\ref{fig:elecsetup}c).
Different configurations for single (I) and double (II~a,b) demodulation are further described in the following section when needed.

\section{Impedancemetry circuit characterization}\label{sec:circhar}

\begin{figure*}[tb]
    \includegraphics{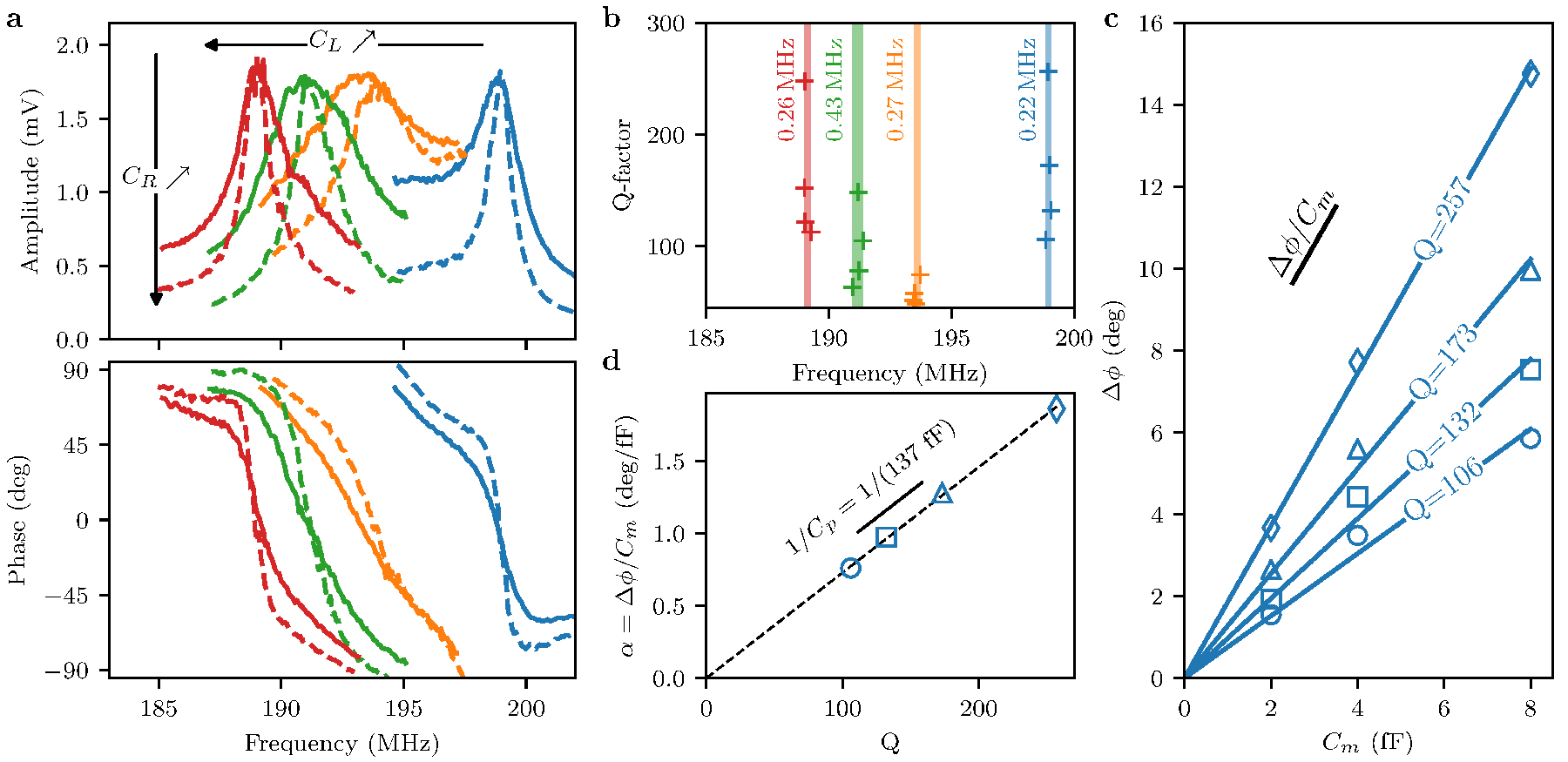}
    \caption{\label{fig:IAres}\textbf{Characterization of the resonant circuit at 4.2~K for capacitance detection. a}, Amplitude and phase of the demodulated circuit output $V_{out}$ for several active inductance settings.
    The resonance frequency shifts to lower frequency as the inductance value increases with increasing $C_L$ (different colors).
    The continuous line (low-$Q$) and dashed line (high-$Q$) show the signals for different values of $C_R$.
    \textbf{b}, Data points for the resonance frequency $f_r$ when the extracted $Q$ is tuned with $C_R$. 
    The colored bars of width given by the written maximal deviation indicate the low dispersion of $f_r$ for fixed $C_L$ when varying $Q$ with $C_R$.
\textbf{c}, Measured phase shift for MOM capacitor $C_m$ of 2, 4, and \SI{8}{\femto\farad} in several $Q$-factor settings.
The capacitance sensitivity $\Delta\phi/C_m$ of the circuit is extracted from the slope with a least square fit at given $Q$.
\textbf{d}, Capacitance sensitivity extracted from \textbf{c} as a function of the $Q$ factor.
A least square linear fit of $\Delta\phi/C_m(Q)$ allows to extract the capacitance $C_p$ parallel with the active inductance.}
\end{figure*}

Without the assistance of low-temperature models, the operating point of the circuit had to be determined experimentally starting from room-temperature settings of bias voltages and currents.
The increase in threshold voltage of NMOS (resp. PMOS) transistors at \SI{4.2}{\kelvin} is compensated by applying a back-gate voltage of \SI{1.2}{\volt} (resp. \SI{-2}{\volt}).
The optimal cryogenic common-mode voltage $V_{cm}=\SI{0.48}{\volt}$ was obtained while monitoring the tank impedance via repeated frequency sweeps until a typical resonance behavior up to \SI{200}{\mega\hertz} emerges for the lowest values of $C_L$ and $C_R$. 
The gain of the low-temperature amplification stage at $f_r$ is optimized with respect to the curent bias of amplifier and buffer (see Supplementary Material~IV).
The main results of the impedancemetry with respect to tunability and detection sensitivity are shown in Figure~\ref{fig:IAres}.

The amplitude and phase of $V_{out}$ using single homodyne detection (I) without any connected DUT are shown in Figure~\ref{fig:IAres}a for the 4 $C_L$ values from 362 to \SI{566}{\femto\farad} and two $C_R$ values chosen between 0 and \SI{322}{\femto\farad} depending on $C_L$.
$V_{out}$ at maximal amplitude was kept equal to \SI{1.8}{\milli\volt} by adjusting $V_{in}$ to avoid non-linearities coming from non-linear MOSFETs behavior.
The resonance frequency $f_r$  varies by \SI{5.1}{\percent} from 189.1 to \SI{199.0}{\mega\hertz} by tuning $C_L$.
The quality factors $Q$ extracted from a linear fit of the phase around $f_r$ are shown in Figure~\ref{fig:IAres}b.
The $Q$ values range from 80 to 250, and can be tuned by a factor $>2$ for every $C_L$ by adjusting $C_R$.
These data demonstrate that $Q$ can be tuned almost independently of the resonance frequency with a frequency variation of less than \SI{0.22}{\percent} across the entire $C_R$ range (see Figure~\ref{fig:IAres}b).

For the minimum value of $C_L$ with the highest resonance frequency, we calibrate the capacitance sensitivity of the circuit for each $Q$ by switching on and off the DUT MOM capacitors $C_m$=2, 3, and \SI{8}{\femto\farad} and  using double homodyne detection (II~a) (see Figure~\ref{fig:IAres}c).
The capacitance sensitivity $\alpha$ is extracted from a least-square linear fit of the phase change $\Delta\phi=QC_m/C_p\equiv \alpha C_m$ for a given $Q$  as shown in Figure~\ref{fig:IAres}d.
The sensitivity $\alpha$ increases linearly with $Q$ from 0.76 to \SI{1.9}{\degree/\femto\farad}.
From the linear fit in Figure~\ref{fig:IAres}d, we obtain $C_p=\SI{137}{\femto\farad}$, in good agreement with the designed value (\SI{136}{\femto\farad}).
In usual circuits without an additional input capacitance\cite{xiao_54-ghz_2007}, the parasitic capacitance of the MOSFETs determines the resonance frequency.
In future design with accurate cryogenic compact models, this capacitance can be reduced significantly leading to higher resonance frequency and improved sensitivity.

From $C_p$ and $f_r$, we are now able to deduce the inductance value $L$.
By adjusting $C_L$, $L$ varies from 2.42 to \SI{5.18}{\micro\henry}.
For a total footprint of \SI{60x50}{\micro\meter}, the active inductance density of \SI{1.73}{\milli\henry/\milli\meter^2} is five orders of magnitude higher than previously used passive inductors (\SI{55}{\nano\henry/\milli\meter^2})\cite{crippa_gate-reflectometry_2019} and three orders of magnitude higher than superconding inductors (\SI{1.6}{\micro\henry/\milli\meter^2})\cite{hornibrook_frequency_2014}.

\section{Capacitance resolution}\label{sec:noise}

\begin{figure}[tb]
    \includegraphics{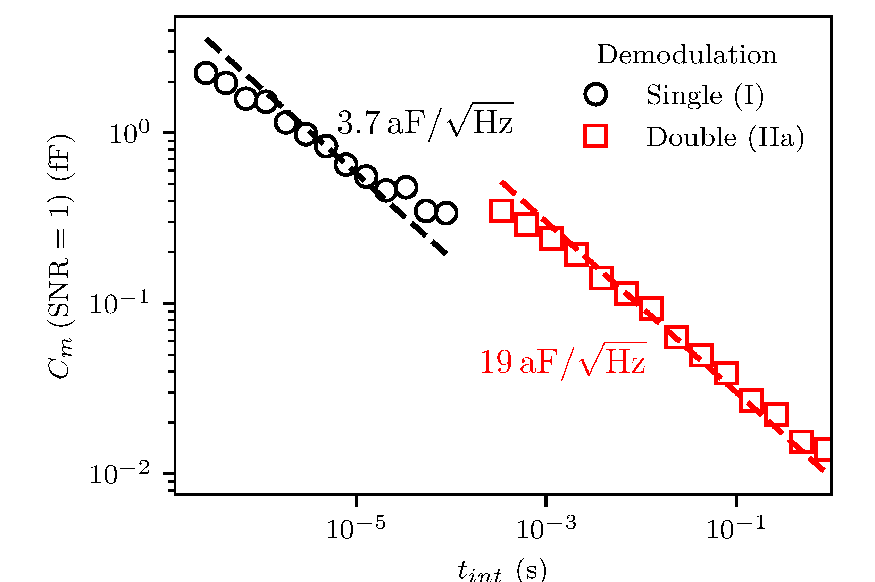}
    \caption{\label{fig:SNR}\textbf{Capacitance resolution of the measurement set-up.} Extrapolated capacitance $C_m$ at signal-to-noise ratio equal to 1 for single (I) (black circles) and double (IIa) (red squares) homodyne detection of the capacitance measurement as a function of the integration time $t_{int}$. 
    Dashed lines are least-square fits $C_m=a\: t_{int}^{-1/2}$ with $a=\sqrt{0.250}\:S_c$ and $S_c$ the equivalent noise spectral density in \si{\atto\farad/\sqrt{Hz}} of the capacitance measurement.}
\end{figure}

We now turn to the resolution in capacitance of the set-up, we derive the input-referred noise in \si{\atto\farad/\sqrt{\hertz}} from the signal-to-noise ratio (SNR) as a function of the integration time $t_{int}$.

For this, we generate a capacitance change by continuously connecting and disconnecting $C_m=\SI{2}{\femto\farad}$ at a rate of \SI{1}{\kilo\hertz}.
Using the demodulation method (I), the square-wave of the phase $\phi$ at $f_r$ with a rise time given by the integration time is used to extract the signal power $P_{sig}$ and noise power $P_{noise}$ by separating the corresponding frequency components in the power spectrum (see Supplementary Material~IV).
The resulting SNR$=P_{sig}/P_{noise}$ is used to extract the capacitance resolution given by the equivalent $C_m(SNR=1)=C_m/SNR$ shown in Figure~\ref{fig:SNR} as a function of $t_{int}$ from \SI{100}{\nano\second} to \SI{100}{\micro\second}. 
A capacitance of \SI{1}{\femto\farad} can be detected with an integration time of \SI{1}{\micro\second} with $SNR=1$.
The capacitance resolution follows a square-root law with $t_{int}$ from which we extract the equivalent input-referred noise of \SI{3.7}{\atto\farad/\sqrt{\hertz}}, two orders of magnitude higher than the best reported sensitivity using an ultra-low noise SQUID amplifier\cite{schupp_radio-frequency_2020}.

As correlated noise appears on time scales longer than $\SI{1}{\milli\second}$ originating probably from the $1/f$ flicker noise of the transistors, we add a second demodulation (IIa) at the capacitance switching frequency of \SI{1}{\kilo\hertz} to remove phase noise originating from a varying $L$.
The \SI{1}{\kilo\hertz} square-wave $\phi$ from (I) with an integration time of \SI{100}{\micro\second} is demodulated by (IIa) at \SI{1}{\kilo\hertz} to obtain its amplitude $|\phi|$.
The capacitance resolution as a function of the second integration time for the \SI{1}{\kilo\hertz} demodulation is extracted by taking the ratio of the average and the standard deviation of the $|\phi|$ signal and is shown in Figure~\ref{fig:SNR}.
With an integration time of \SI{1}{\second}, the resolution becomes as low as \SI{10}{\atto\farad}.

\section{Quantum capacitance measurements}\label{sec:qcapa}

\begin{figure*}[tb]
    \includegraphics{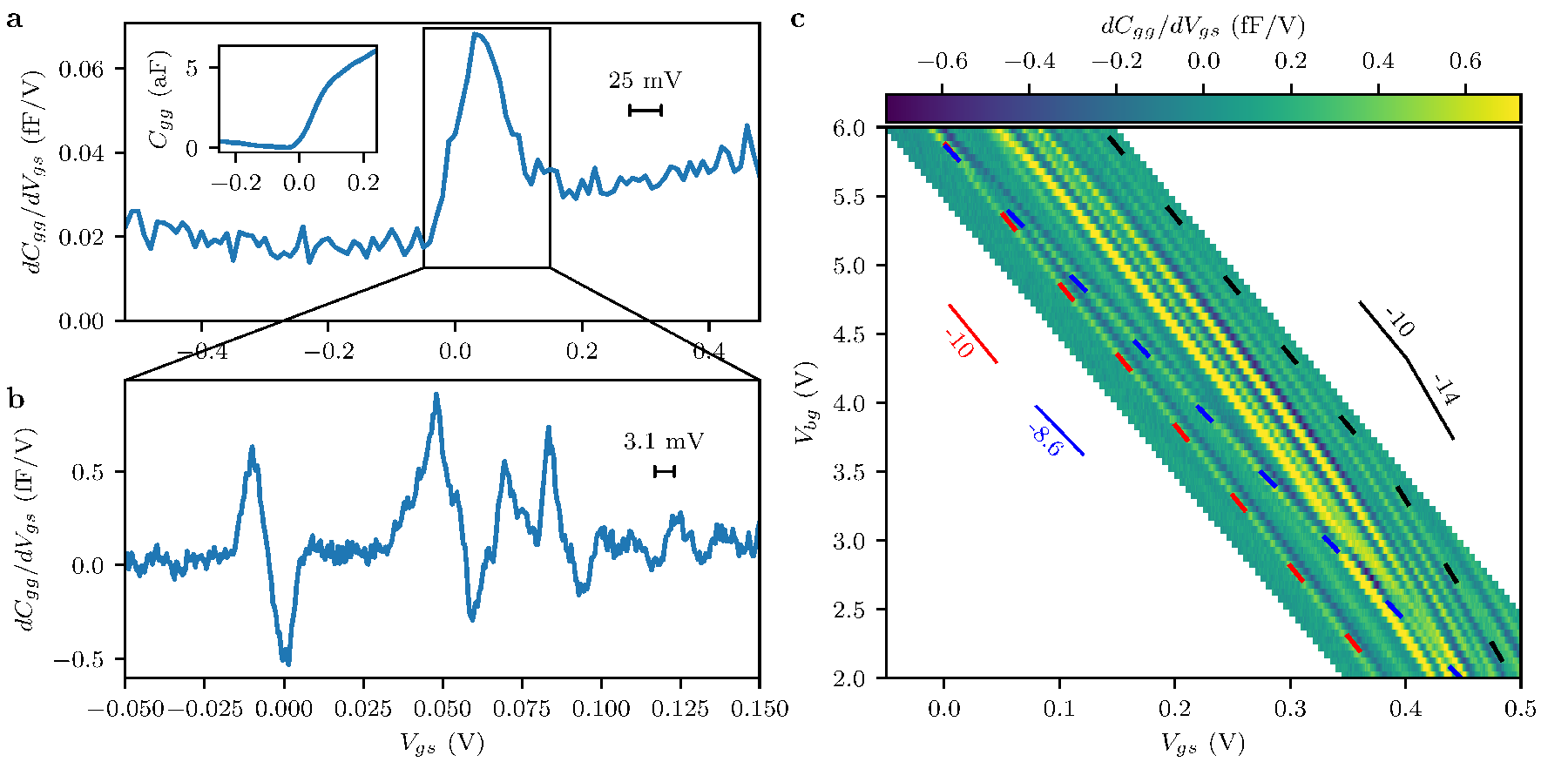}
    \caption{\label{fig:qcapa}\textbf{Quantum capacitance measurement of an integrated MOSFET with channel length 120~nm and width 80~ nm. a}, Measurement of the first derivative of the gate capacitance $C_{gg}$ with respect to $V_{gs}$ by applying a gate-source AC excitation of \SI{25}{\milli\volt}.
        The inset shows the capacitance $C_{gg}(V_{gs})$ computed from the integrated signal of the derivative.
        \textbf{b}, Expanded view of $dC_{gg}/dV_{gs}$ around the off-on transition of the MOSFET measured with a smaller excitation of \SI{3.1}{\milli\volt}.
        The resolved features are signatures of quantized electronic states in the measured capacitance of the MOSFET channel.
    \textbf{c}, Evolution of $dC_{gg}/dV_{gs}$ with the back-gate voltage $V_{bg}$ and the gate-source voltage $V_{gs}$.
    The indicated slopes $\beta=dV_{bg}/dV_{gs}\simeq C_{g-ch}/C_{bg-ch}$ represent the relative coupling strength of the detected quantized states with respect to back- and front-gate. 
}
\end{figure*}

With the calibrated impedancemetry circuit, we are able to detect the gate quantum capacitance $C_{gg}$ of the multiplexed tiny MOSFETs (M0, M1, M2 in Figure~\ref{fig:elecsetup}a) similar to the ones used to implement spin qubits\cite{maurand_cmos_2016} or single-electron transistors for read-out\cite{} with CMOS technology.
Measurements will be presented for M2 with a gate length of \SI{120}{\nano\meter} and a gate width of \SI{80}{\nano\meter}.
The other available device shows similar behavior (see Supplementary Material~V).

The total gate capacitance $C_{gg}$ of such devices corresponds to the sum of the capacitance to drain, source, back-gate, and MOSFET channel of which the gate to channel capacitance depends highly on the gate-source voltage $V_{gs}$ controlled by the DC component of $V_{bias}$ (see Figure~\ref{fig:elecsetup}a).
As $C_{gg}$ of nanometric devices is extremely small compared to $C_p$,  we don't expect to have sufficient SNR for small capacitance variations at reasonable integration times.
Better sensitivity can be obtained by modulating $V_{gs}$ (method IIb) to measure after demodulation the first derivative $dC_{gg}/dV_{gs}$ as $C_{gg}$ varies a lot in a small $V_{gs}$ window. 

While the resonator impedance is probed at \SI{199}{\mega\hertz}, $V_{gs}$ is modulated at \SI{1}{\kilo\hertz} with mV-range excitation on $V_{bias}$ (see Figure~\ref{fig:elecsetup}bc).
The obtained result with a relatively large \SI{25}{\milli\volt} $V_{gs}$ modulation (shown in Figure~\ref{fig:qcapa}a) is reminiscent of the typical gate capacitance variation around threshold voltage $V_{th}\simeq\SI{0}{\volt}$ at $V_{bg}=\SI{6}{\volt}$\cite{}.
$C_{gg}$ (see inset of Figure~\ref{fig:qcapa}a) reflects the typical behavior for a FET capacitance from the subthreshold regime $V_{gs}\ll V_{th}$ to the strong inversion regime $V_{gs}\gg V_{th}$. 
Upon decreasing the amplitude of the $V_{gs}$ modulation to only \SI{3.1}{\milli\volt}, the observed $dC_{gg}/dV_{gs}$ signal in Figure~\ref{fig:qcapa}b reveals a fine structure around $V_{th}$ consisting of successive peak-dip oscillations.
Following numerical integration, these features result in a series of peaks in $C_{gg}$, which we interpret as quantum contributions to the capacitance coming from electrons tunneling in and out of localized quantum states within the transistor channel.

To further identify these quantum states, we acquire $dC_{gg}/dV_{gs}$ for different back-gate voltage $V_{bg}$ from 2 to \SI{6}{\volt} as shown in Figure~\ref{fig:qcapa}c.
As $V_{bg}$ increases, all observed features shift to lower $V_{gs}$  with a slope close to the ratio $\beta$
of gate-channel capacitance $C_{g-ch}$ over the backgate-channel capacitance $C_{bg-ch}$ alike the $V_{th}$-shift with back-gate for similar FET devices\cite{paz_variability_2020}.
For $V_{bg}>\SI{4}{\volt}$, all features have a coupling ratio of 10 except for one with a lower coupling 8.6 attributed to an impurity closer to the back-gate interface.
No anomalous impurity structure is detected in the smaller \SI{60x80}{\nano\meter} device (see Supplementary Material~V).
At lower $V_{bg}$, the coupling increases with $V_{gs}$ from 10 to 14 as the electron-filled inversion layer is brought back to the top-interface.

These measurements of integrated quantum devices demonstrate that the capacitive signature of structure in the electronic density of states of quantum dots can be probed via impedancemetry.

\section{Conclusions}
We reported an integrated circuit that performs impedancemetry of a resonator coupled to a quantum dot at cryogenic temperatures.
The active inductance of the resonator allowed the controlled tuning of the resonance frequency and quality factor, which will be of importance for optimal frequency-spectrum crowding in multiplexed read-out schemes.
The employed multiplexing of nanometric quantum devices with on-chip switches could be beneficial for reduced power per qubit in a scalable multi-qubit architecture.
Novel read-out architectures with cryogenic electronics, such as the active inductance, have the potential to increase scalability and flexibility in the design and exploitation of quantum processors.

Further work towards lower noise and lower power design with more accurate high-frequency models at cryogenic temperatures will improve the final performance.
Measuring multiplexed out-of-chip capacitances of quantum devices will be also promising for the screening of quantum devices with a simpler experimental setup than reflectometry.
In the long run, the realization of tailored high-end analog electronics at cryogenic temperatures  will improve and accelerate the up-scaling of quantum processors.

\section*{Methods}
\textbf{Fabrication details.}
The impedancemetry chip was designed in a commercial CMOS FD-SOI 28nm technology with low-$V_{th}$ (LVT) thin-oxide (GO1) transistors.
The chip is wire-bonded onto a QFN48 package directly soldered on a 4-layer PCB with FR4 substrate.

\textbf{Measurement set-up.}
The FR4 PCB is placed at the end of a dip-stick enclosed in a metallic container filled with a small amount of helium gas for thermal exchange with a liquid He bath (see Supplementary Material~VI).
A PCB-mounted thermometer ensures a precise monitoring of the PCB temperature. 
High-frequency lines of $V_{in}$ and $V_{out}$ are routed on the PCB from the chip package to the SMA coaxial connectors via top-layer 50$\Omega$-matched coplanar waveguide with ground plane and via fencing.
Supply lines are decoupled from environmental noise with PCB-mounted capacitors (0.1, 1, \SI{10}{\micro\farad}) and conveyed to room temperature with copper wiring.
All other DC lines are conveyed to room-temperature with $50-\SI{130}{\ohm}$ constantan wiring.
At room temperature, electronic apparatus comprises a multi-channel low-noise 21-bit digital-to-analog converter, and a 600~MHz lock-in amplifier.

\section*{Data availability}
The data that support the plots within this paper and other findings of this study are available from the corresponding author upon reasonable request.

\bibliographystyle{unsrt}
\bibliography{ms.bbl}
\section*{Acknowledgements}
This work was partly supported by the European Union’s Horizon 2020 research and innovation program under Grant Agreement No. 810504 (ERC Synergy project QuCube).

\section*{Competing interests}
The authors declare a patent application FR1914651, filed on december 17th 2019.

\end{document}


\title{Supplementary Material: Compact gate-based read-out of multiplexed quantum devices with a cryogenic CMOS active inductor}

\author{L.~Le~Guevel}
\email{loick.leguevel@gmail.com}
\affiliation{Univ. Grenoble Alpes, CEA, LETI, F-38000 Grenoble, France}
\affiliation{Univ. Grenoble Alpes, CEA, Grenoble INP, IRIG, PHELIQS, F-38000 Grenoble, France}
\author{G.~Billiot}
\affiliation{Univ. Grenoble Alpes, CEA, LETI, F-38000 Grenoble, France}
\author{S.~De~Franceschi}
\affiliation{Univ. Grenoble Alpes, CEA, Grenoble INP, IRIG, PHELIQS, F-38000 Grenoble, France}
\author{A.~Morel}
\affiliation{Univ. Grenoble Alpes, CEA, LETI, F-38000 Grenoble, France}
\author{X. Jehl}
\author{A.G.M.~Jansen}
\affiliation{Univ. Grenoble Alpes, CEA, Grenoble INP, IRIG, PHELIQS, F-38000 Grenoble, France}
\author{G.~Pillonnet}
\affiliation{Univ. Grenoble Alpes, CEA, LETI, F-38000 Grenoble, France}

\date{\today}
             
\maketitle
\onecolumngrid

\FloatBarrier
\section{Read-out scaling of large qubit arrays}
\FloatBarrier

\begin{figure}[h]
\begin{tikzpicture}
    \node[anchor=south west, inner sep=0pt] (image) at (0,0)
    {\includegraphics{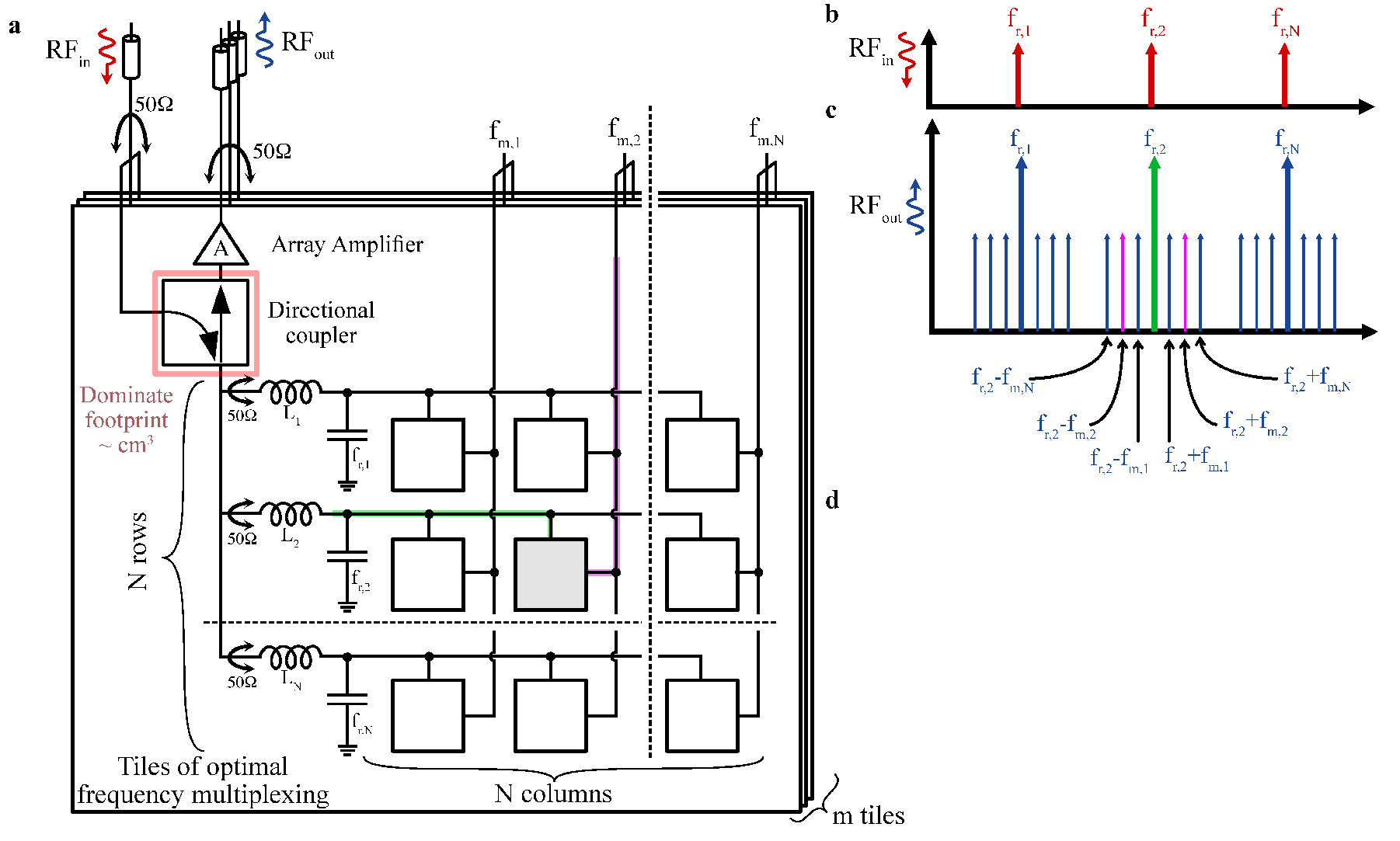}};
    \begin{scope}[x={(image.south east)},y={(image.north west)}]
        \node[fill=white, anchor=south east] at (0.97,0.06)
        {\begin{tabularx}{6cm}{
   >{\centering\arraybackslash}X 
   >{\centering\arraybackslash}X 
   >{\centering\arraybackslash}X}  
            \toprule
            
            Element & Footprint &  Scaling \\
            & (mm$^2$) &   \\
            \midrule
            Qubits & $10^{-8}$ & $m\times N^2$\\
            Amplifiers & $0.1$ &  $m$\\
            Inductors & $1$ & $m\times N$\\
            Couplers & $100$ &  $m$ \\
            \bottomrule
    \end{tabularx}};
    \end{scope}
\end{tikzpicture}
    \caption{\textbf{Reflectometry and scaling of large qubit arrays using directionnal couplers and passive inductors.}\label{figSM:scaleReflecto} \textbf{a}, Circuitry for simultaneous read-out of large arrays of qubits using reflectometry  with frequency multiplexing.
        To address $N\times N$ qubits in an array, $N$ resonators at different resonant frequencies $f_{r,i}$ are each coupled to a row of $N$ qubits frequency-multiplexed with $f_{m,j}$ .
        Resonators are probed with the scattering parameter $S_{11}$ by sending a voltage excitation made of a frequency-comb $f_{r,i}$.
    The incoming wave $RF_{in}$ and outgoing $RF_{out}$ wave are separated with a directional coupler.
    The resonator frequency is defined by the $\SI{50}{\ohm}$ matching of the equivalent impedance of all resonators and differs from the LC resonant frequencies.
    $RF_{out}$ is sent to room-temperature demodulation with an amplifier.
    Typical frequency-spectrum crowding is represented in \textbf{b-c}.
    Frequency multiplexing for a given finite bandwidth and given read-out time imposes an upper-bound on the array size $N$.
    As an example, for a read-out time of \SI{1}{\micro\second} and bandwidth of \SI{1}{\giga\hertz}, $N\times N$ can only be as high as $100$.
    To further increase the number of qubits, each tile containing $N\times N$ qubits with the  required circuitry is duplicated $m$ times.
    The input frequency-comb voltage excitation $RF_{in}$ is common to all tiles while the output signal $RF_{out}$ requires for each of them a coaxial line.
\textbf{d}, Scaling  laws in $m$ and $N$ and the typical footprint for the circuit elements.
The footprint of reflectometry is limited by the directional coupler size of $\sim\SI{1}{\centi\meter^2}$ (red square in \textbf{a}). }
\end{figure}

\begin{figure}[h]
    
\begin{tikzpicture}
    \node[anchor=south west, inner sep=0pt] (image) at (0,0)
    {\includegraphics{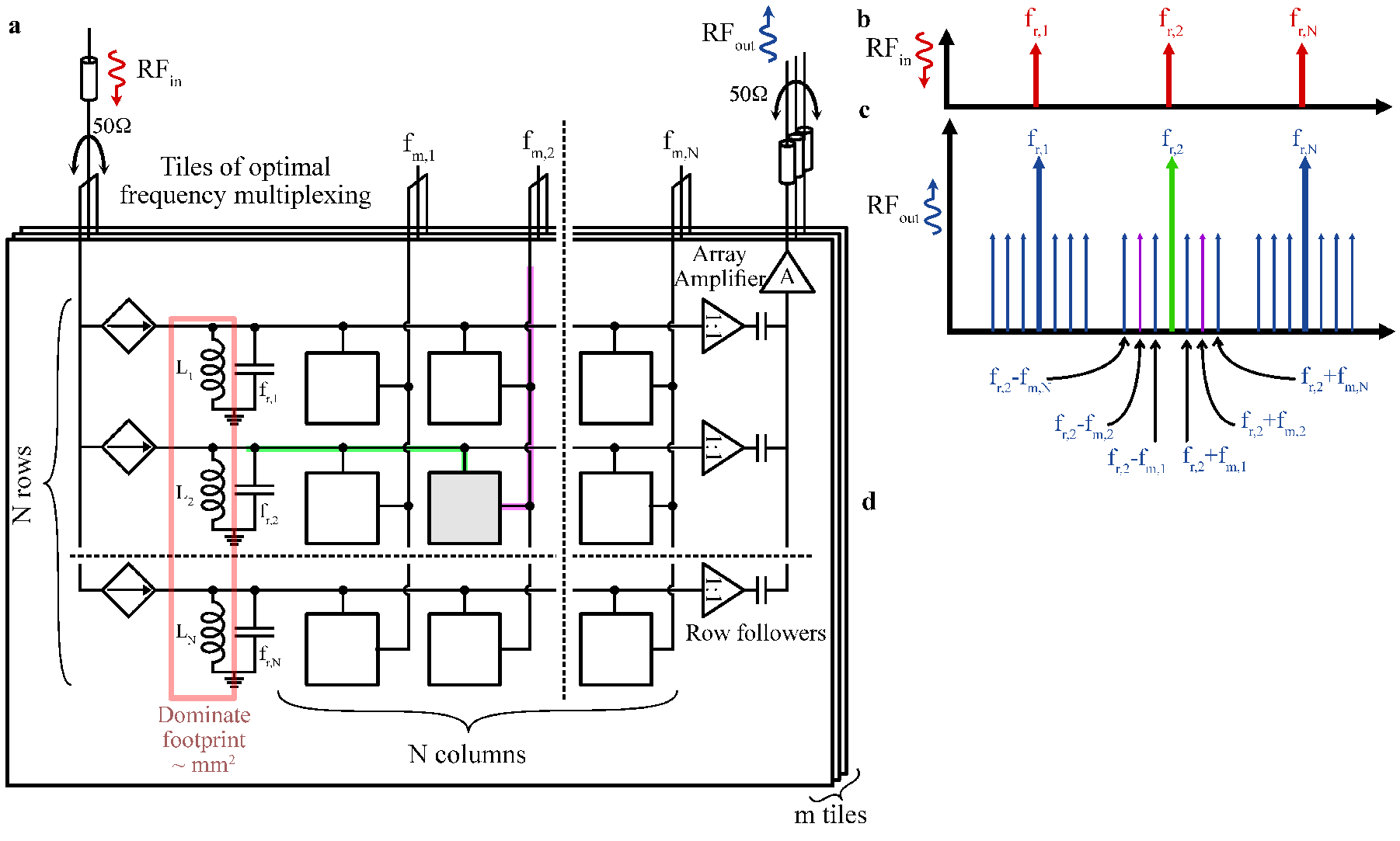}};
    \begin{scope}[x={(image.south east)},y={(image.north west)}]
        \node[fill=white, anchor=south east] at (0.97,0.06)
        {\begin{tabularx}{6cm}{
   >{\centering\arraybackslash}X 
   >{\centering\arraybackslash}X 
   >{\centering\arraybackslash}X}  
            \toprule
            
            Element & Footprint &  Scaling \\
            & (mm$^2$) &   \\
            \midrule
            Qubits & $10^{-8}$ & $m\times N^2$\\
             Source &  $10^{-3}$ &  $m\times N$\\
            Followers & $10^{-3}$ &  $m\times N$\\
            Amplifiers & $0.1$ &  $m$\\
            Inductors & $1$ & $m\times N$\\
            \bottomrule
    \end{tabularx}};
    \end{scope}
\end{tikzpicture}
\caption{\textbf{Impedancemetry and scaling of large qubit arrays using passive inductors.}\label{figSM:scale} \textbf{a}, Circuitry for simultaneous read-out of large arrays of qubits using impedancemetry with frequency multiplexing.
    To address $N\times N$ qubits in an array, $N$ resonators at different resonant frequencies $f_{r,i}$ are each coupled to a row of $N$ qubits frequency-multiplexed at $f_{m,j}$  for current excitation via voltage-controlled current sources. 
    Each resonator filters the out-of-resonance components within the frequency comb $f_{r,i}$ of input excitation.
    The voltage responses of each resonator are added together and sent through an amplifier to room-temperature demodulation for signal recovery of every single qubit response.
    Typical frequency-spectrum crowding is represented in \textbf{b-c} for the input and output signals.
    Frequency multiplexing for a given finite bandwidth and given read-out time imposes an upper-bound on the array size $N$.
    As an example, for a read-out time of \SI{1}{\micro\second} and bandwidth of \SI{1}{\giga\hertz}, $N\times N$ can only be  as high as $100$.
    To further increase the number of qubits, each tile containing $N\times N$ qubits  is duplicated $m$ times.
    The input frequency-comb voltage excitation $RF_{in}$ is common to all tiles while the output signal $RF_{out}$ requires  for each of them a coaxial line.
    \textbf{d}, Scaling laws in $m$ and $N$ and the  typical footprint for the circuit elements. The footprint of impedancemetry  is limited by the inductor size (red square in \textbf{a}). Passive inductors with footprint of \SI{1}{\milli\meter^2} can be replaced by controllable active inductors with a footprint of only \SI{0.001}{\milli\meter^2} to improve the circuitry scalability.
    }
\end{figure}

\FloatBarrier
\clearpage
\section{Design of the integrated circuit}
\FloatBarrier

\begin{figure}[h]
    \includegraphics{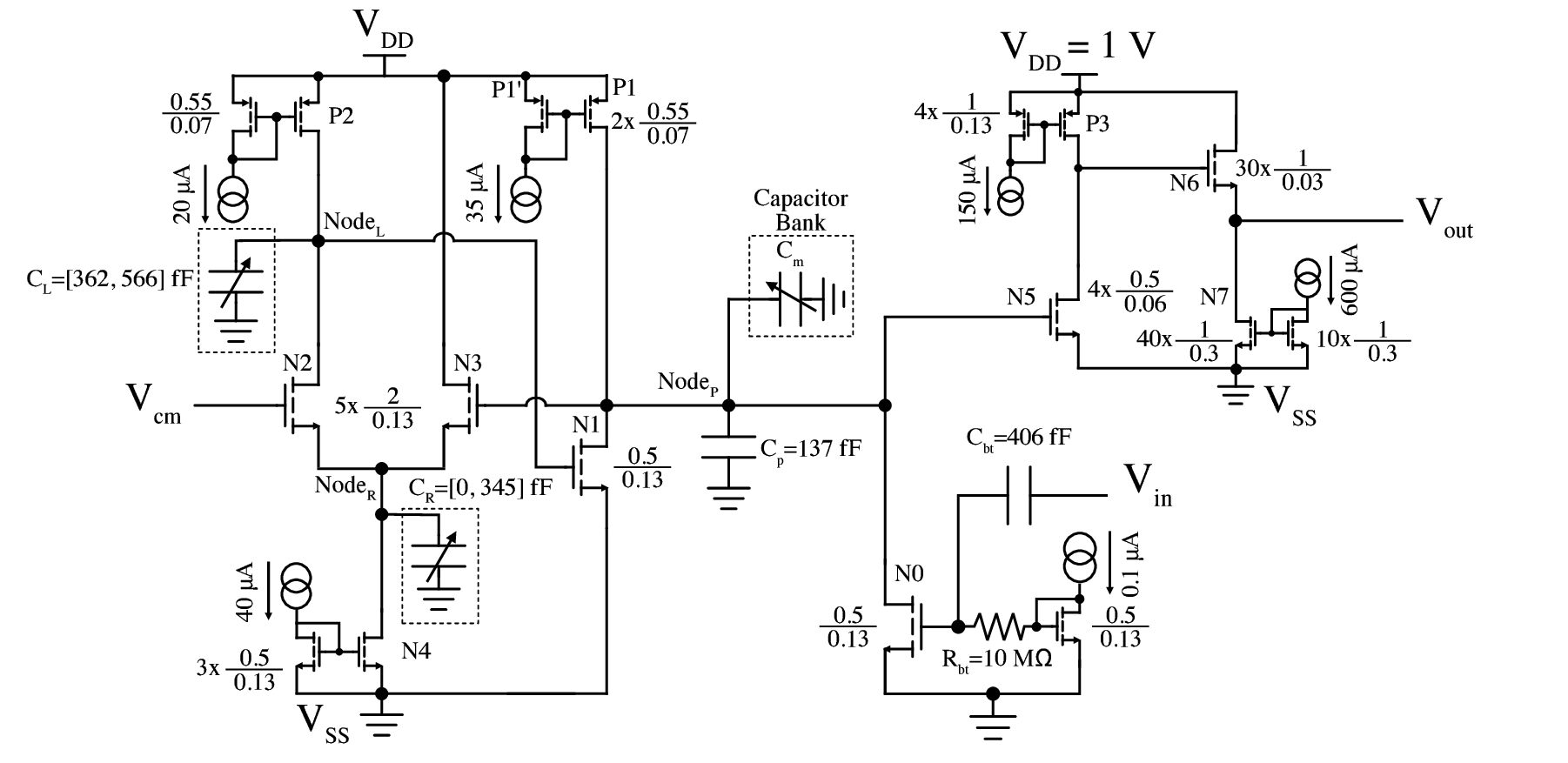}
    \caption{\label{figSM:implem}\textbf{Design implementation of the impedancemetry chip.}  Details of the components with transistor dimensions in the integrated circuit.
        The capacitor banks used as variable capacitors ($C_L$, $C_R$, and $C_m$) are detailed in Figure~\ref{figSM:tunablecap}~and~\ref{figSM:DUTmux}. 
        Each transistor dimension is indicated as $m\times\frac{W}{L}$ with $W$ (resp. $L$) the gate-finger  width (resp. length) of the gate in \si{\micro\meter} and $m$ is the number of fingers.
        The indicated current references of the  diode-mounted transistors are generated at room temperature (see Figure~\ref{figSM:expsetup}).
    }
\end{figure}

\begin{figure}[h]
    \includegraphics{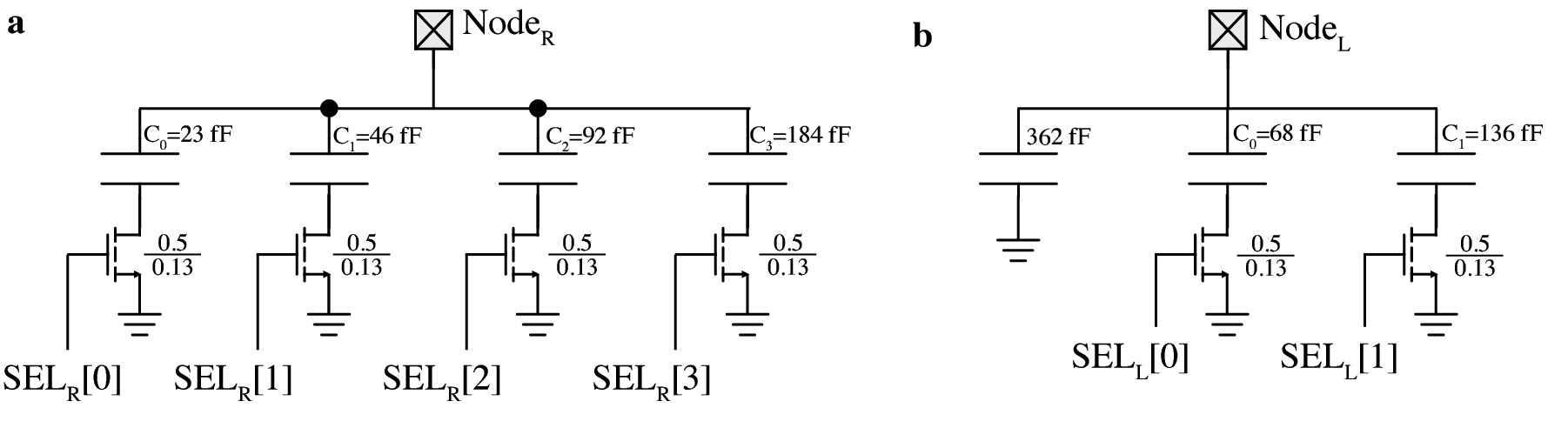}
    \caption{\textbf{Design implementation of the variable capacitors.} \textbf{a}, Implementation of the variable capacitor $C_R$ at Node$_R$ in Figure~\ref{figSM:implem}.  The 4 binary-weighted MOM capacitors are selected with NMOS switches activated by the selection bits $SEL_R[3:0]$ for $C_R$ variation from 0 to 345~fF.
    \textbf{b}, Similar implementation of $C_L$ with 3 binary-weighted MOM capacitors in parallel from 362 to 566~fF. \label{figSM:tunablecap}}
\end{figure}

\begin{figure}[h]
    \includegraphics{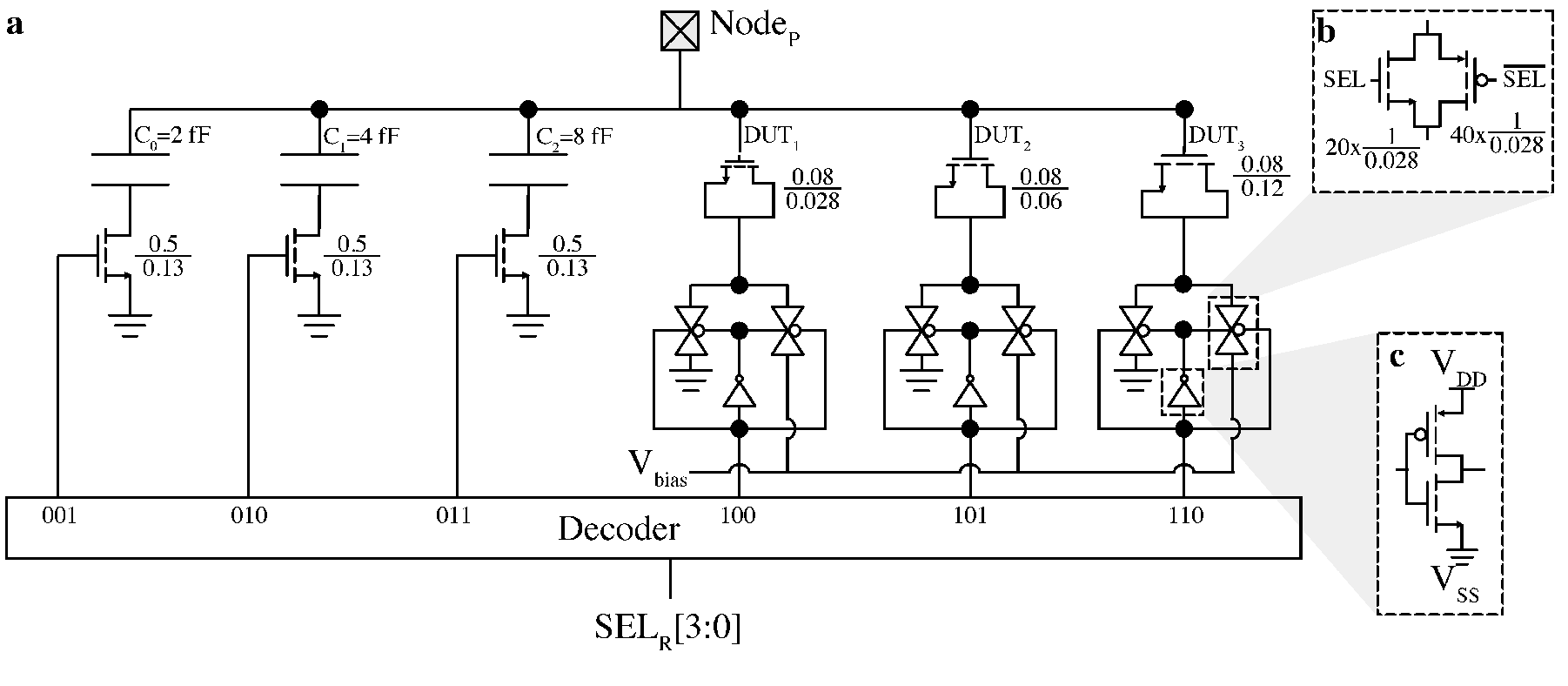}
    \caption{\textbf{Multiplexing of the Device Under Test (DUT).}\label{figSM:DUTmux} \textbf{a}, Design implementation of the capacitor bank with 3 MOM capacitors for capacitive calibration and 3 nanometer-sized  MOSFETs for quantum capacitance measurements. The MOM capacitors are selected with NMOS switches. The MOSFETs are selected with a pair of pass-gates as  detailed in \textbf{b}. 
    When the MOSFETs are unselected, the drain and source are set to ground while the selected-MOSFET drain and source are linked to $V_{bias}$ to change $V_{gs}$.
Pairs of pass-gate are made complementary (when one is OFF, the other one is ON) with one inverter as  shown in \textbf{c}.}
\end{figure}

\begin{figure}[h]
\begin{tikzpicture}
    \node[anchor=south west, inner sep=0pt] (image) at (0,0)
    {\includegraphics{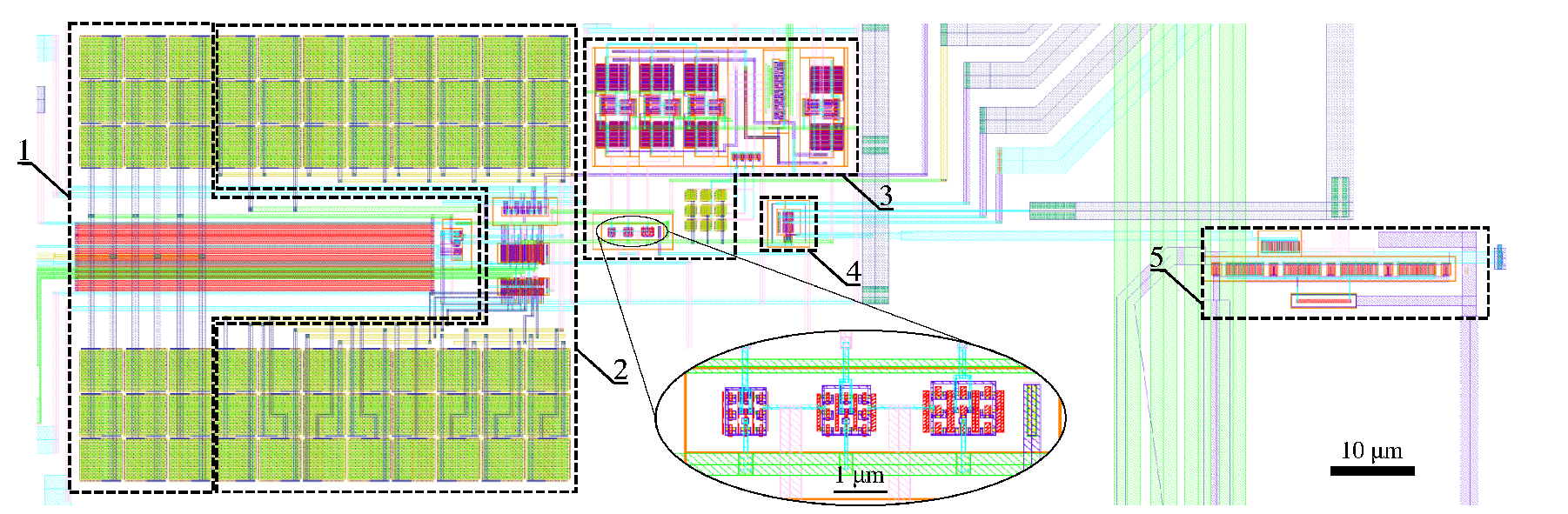}};
    \begin{scope}[x={(image.south east)},y={(image.north west)}]
        \node[fill=white, anchor=north east] at (0.97,1.13)
    {\begin{tabular}{ccS[table-format=1.6]S[table-format=2.2]}
            \toprule
            
            ID & Block name & {Area (\si{\milli\meter^2})} & {Contribution (\si{\percent})} \\
            \midrule
            1 & Current generation & 0.0013 & 33 \\
            2 & Active inductance & 0.0017 & 41 \\
            3 & Capacitor bank & 0.00067 & 17 \\
            4 & Amplifier & 0.000035 & 0.86 \\
            5 & Follower & 0.00035 & 8.7\\
            \midrule
            & Total & 0.0040 & 100\\

            \bottomrule

    \end{tabular}};
    \node[anchor=north west]  at (0,1.13) {\textbf{a}};
    \node[anchor=north west]  at (0.559,1.13) {\textbf{b}};
    \end{scope}
\end{tikzpicture}
\caption{\textbf{Layout and footprint of the impedancemetry circuit.} \textbf{a}, Layout view of the impedancemetry circuit. The labeled areas correspond to the circuit block names of table \textbf{b}.
The bottom inset shows a zoomed window for the measured quantum devices. Each device is surrounded by dummies to improve the fabrication quality of the nanometer-sized  devices. \textbf{b}, Table of the occupied area of each block for a total footprint of \SI{0.004}{\milli\meter^2}.  }
\end{figure}

\newpage
\phantom{l}
\newpage
\phantom{l}

\FloatBarrier
\FloatBarrier
\section{Simulation Results at 300~K}
\FloatBarrier

\begin{figure}[h]
    \includegraphics{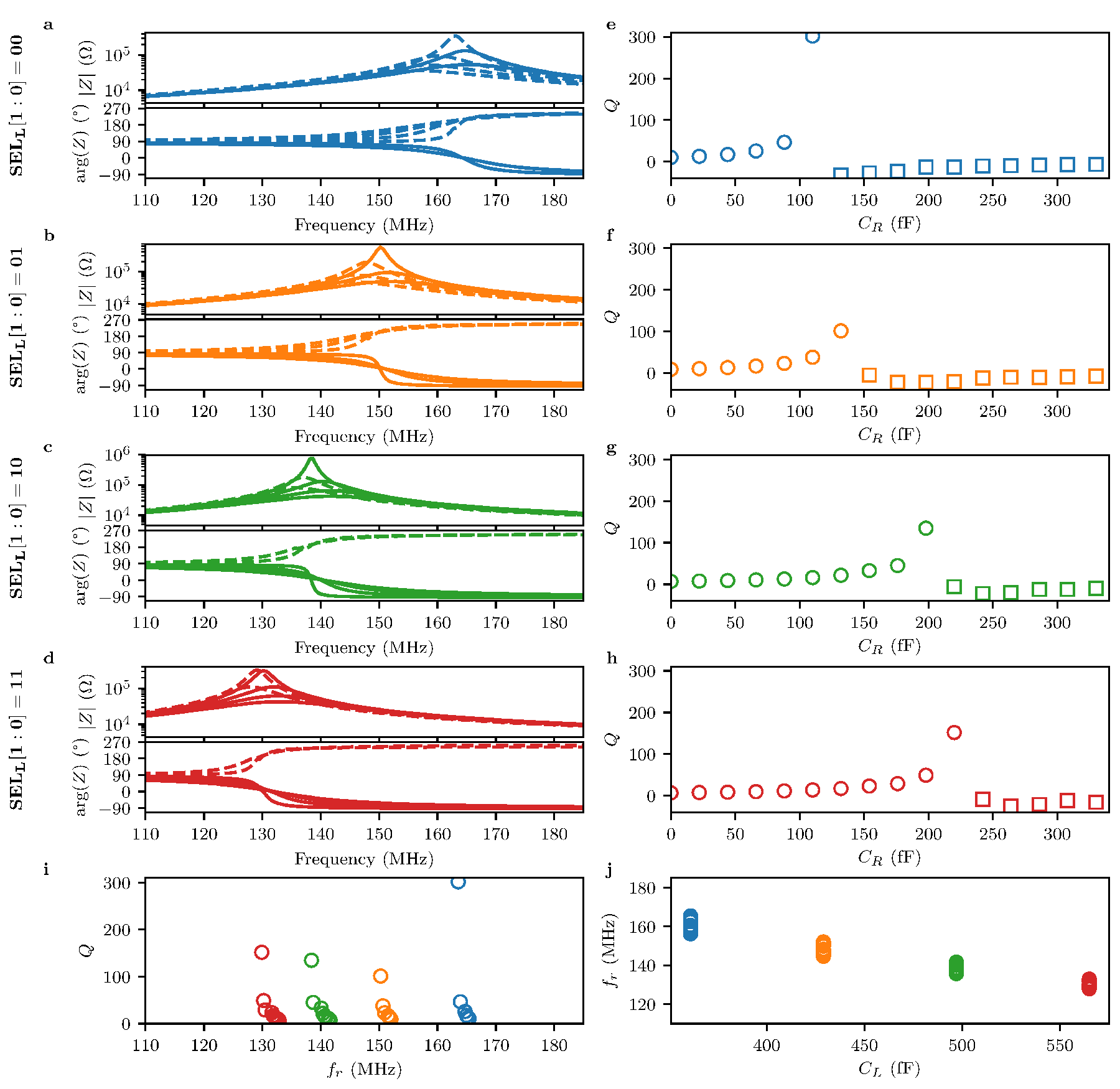}
    \caption{\textbf{Impedance of the active inductance from simulations at 300~K.} \textbf{a-d}, Complex impedance of the active inductance for a few $C_R$ values with $C_L$ equal to: \textbf{a} 362, \textbf{b} 420, \textbf{c} 498, and \textbf{d} \SI{566}{\femto\farad}. \textbf{e-h}, Quality factor $Q$ of the active inductance as a function of $C_R$ extracted from the tank impedance.
    In \textbf{a-d} (respectively \textbf{e-h}), stable resonance data with $Q\ge 0$ are shown in continuous lines (resp. round markers) while unstable resonance data with $Q<0$ are shown as dashed lines (resp. square markers).
\textbf{i}, $Q$ as a function of the resonant frequency $f_r$ showing small dispersion.
\textbf{j}, Evolution of $f_r$ with $C_L$ at all $C_R$ values.}
\end{figure}
\begin{SCfigure}[1][h]
    \includegraphics[scale=0.8]{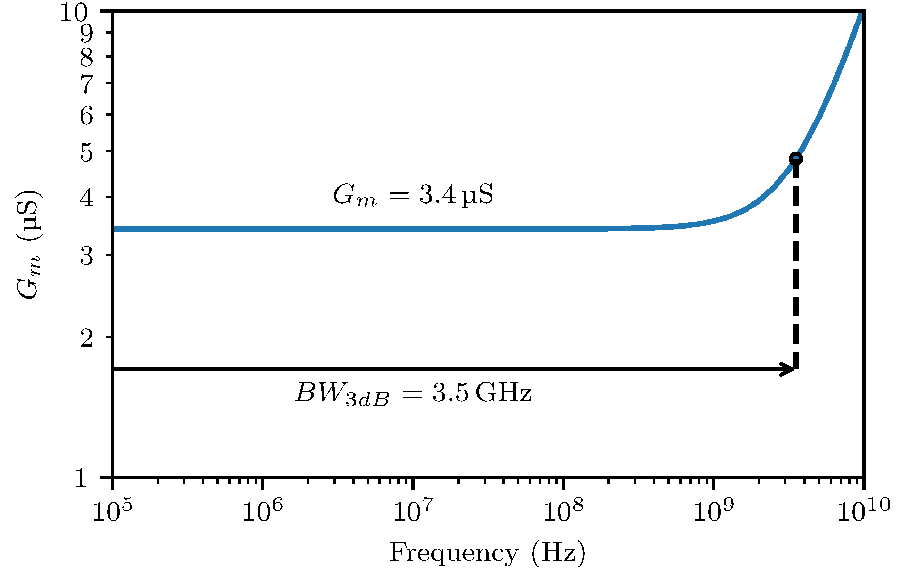}
    \caption{\textbf{Voltage-to-current conversion for the current excitation of the tank from simulations at 300~K.} Transimpedance $G_m$ and bandwidth of the current generation as a function of frequency extracted from foundry models at 300~K.}
\end{SCfigure}
\begin{figure}[h]
    \includegraphics{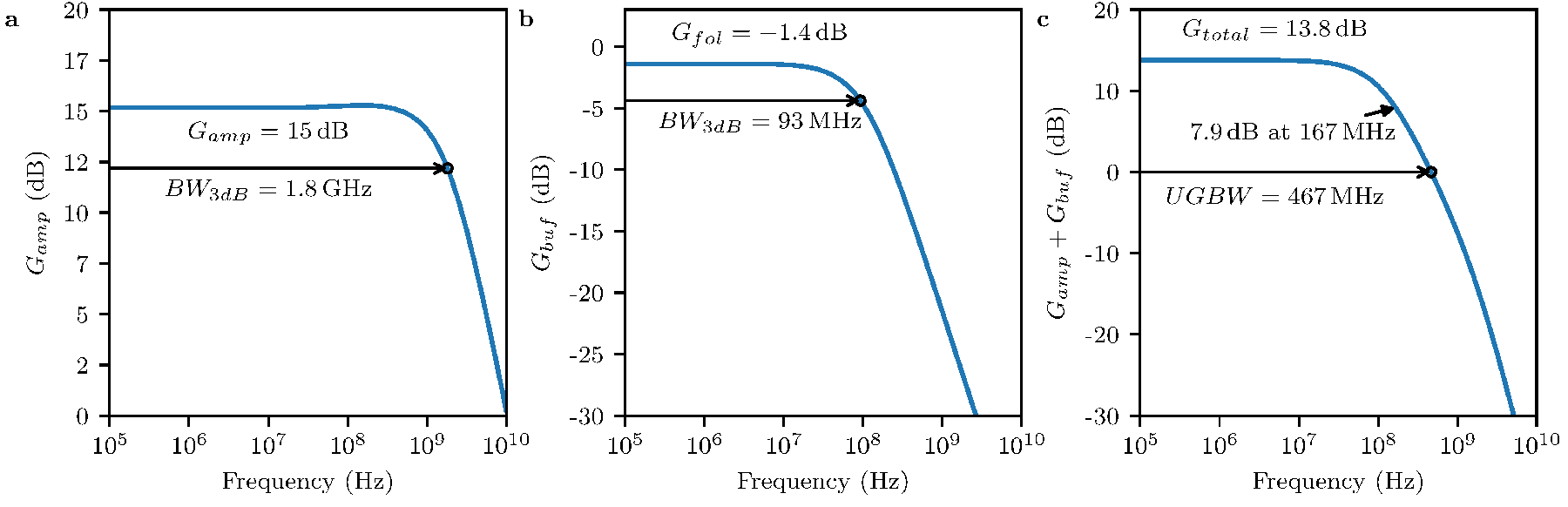}
    \caption{\textbf{Amplifier and follower characteristics from simulations at 300~K.} \textbf{a}, Gain of the amplifier as a function of frequency extracted from AC simulations with foundry models at 300~K. \textbf{b}, Gain of the follower loading a \SI{50}{\pico\farad} cable capacitance at the chip output. \textbf{c}, Total amplification of amplifier and follower as a function of frequency. Despite large cable capacitance from 4.2~K to 300~K stage, the unity-gain bandwidth of \SI{467}{\mega\hertz}
    allows to keep the gain above 1  at the resonant frequencies $f_r$. }
\end{figure}

\begin{SCtable}[2][h]
    \caption{\textbf{Noise contribution from linear AC simulations at 300~K.} Listed noise contributions of the 7 transistors generating \SI{84}{\percent} of the total output noise of the impedancemetry chip extracted from foundry models at 300~K. The listed transistor devices belong to the active inductance which constitutes the main source of noise in the circuit.}. 
    \begin{tabular}{ccc}
        \toprule
        Rank & Device & Contribution\\
        \midrule
        1 & P1  & 25\%\\
        2 & N1 & 23\%\\
        3 & P1'  & 14\%\\
        4 & P2 & 10\%\\
        5 & N4 & 5\%\\
        6 & N2 & 4\%\\
        7 & N3 & 3\%\\

        \bottomrule
    \end{tabular}
\end{SCtable}
\begin{SCfigure}[10][h]
    \includegraphics{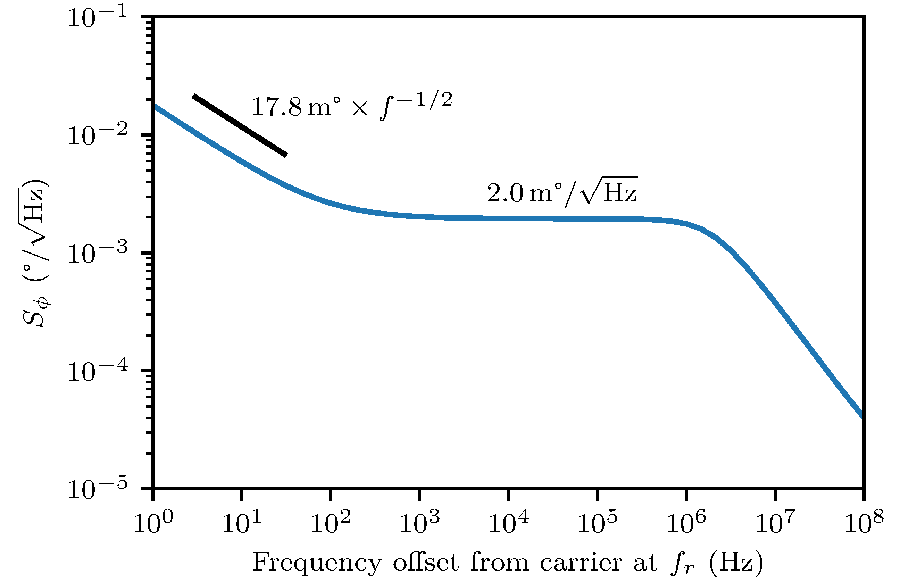}
    \caption{\textbf{Phase noise of the impedancemetry setup from 300~K simulations.} Phase noise as a function of the frequency offset with respect to the resonance frequency of the tank $f_r=\SI{167}{\mega\hertz}$ extracted at the chip output from Steady STate (SST) simulations at 300~K with foundry models. 
        The plateau above about 100~Hz with the  cut-off around $\sim\SI{2}{\mega\hertz}$ is the phase noise behavior we could expect from a linear AC noise analysis.
        The cut-off corresponds to the width of the resonant peak $\kappa\sim f_r/Q\simeq \SI{2.1}{\mega\hertz}$ with the quality factor $Q=81$. 
        From the plateau value of \SI{2}{\milli\degree/\sqrt{\hertz}}, we estimate the input-referred noise to \SI{3.7}{\atto\farad/\sqrt{\hertz}}.
        The appearance of a flicker component in the phase noise is the evidence of the mixing between $1/f$ low-frequency noise and the tank high-frequency signal.
    This additional non-linear noise is attributed to transistor noise that translates into noise contribution to the inductance $L$ and quality factor $Q$ resulting in low-frequency noise up-mixing.}
\end{SCfigure}

\FloatBarrier
\newpage
\section{Complementary data of the resonant circuit at 4.2~K}
\FloatBarrier

\begin{figure}[h]
    \includegraphics{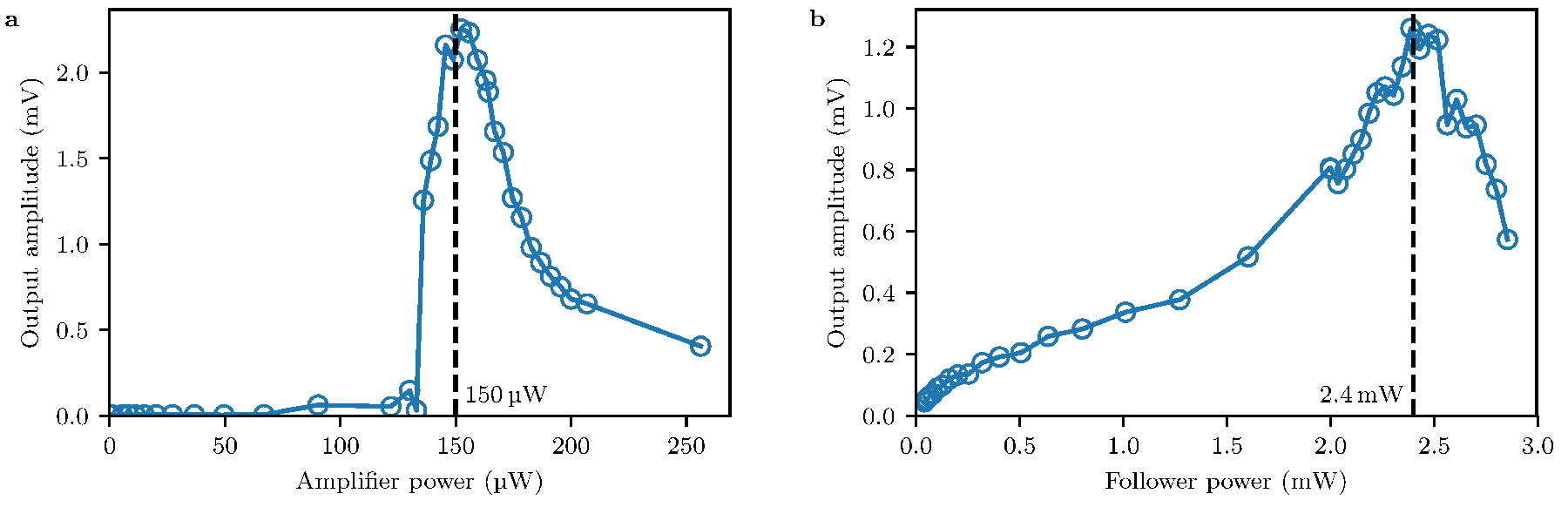}
    \caption{\textbf{Amplifier and follower optimization at 4.2~K.} \textbf{a}, Output voltage $V_{out}$ at the tank resonant frequency $f_r=\SI{199}{\mega\hertz}$ as a function of the amplifier power at 4.2~K.
    To maximize the gain, we choose the operating power of the amplifier at \SI{150}{\micro\watt}.
    \textbf{b}, Output voltage $V_{out}$ at the tank resonant frequency $f_r=\SI{199}{\mega\hertz}$ as a function of the follower power at 4.2~K.
To maximize the gain, we choose the operating power of the follower at \SI{2.4}{\milli\watt}.}
\end{figure}

\begin{figure}[h]
    \includegraphics{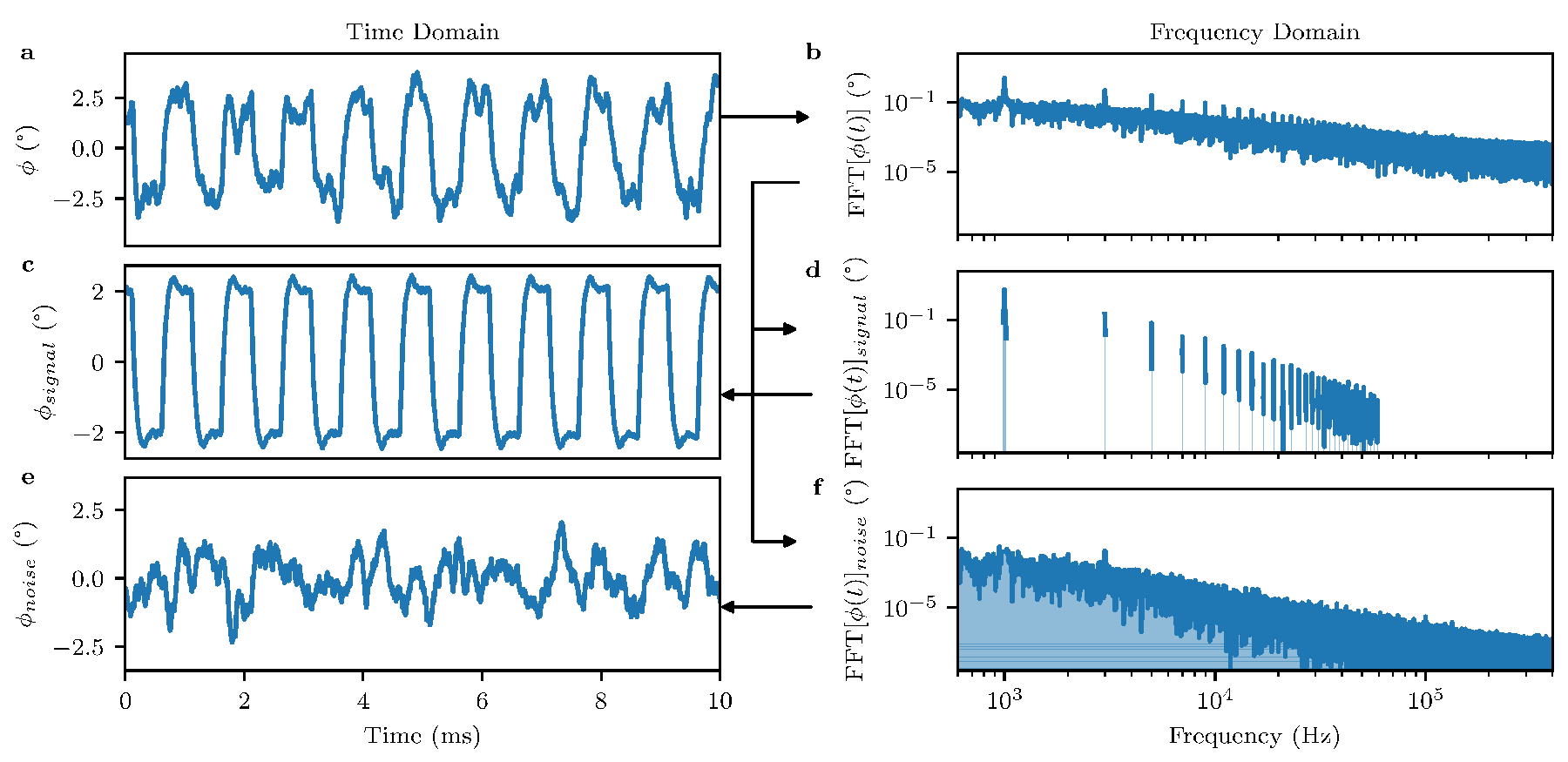}
    \caption{\textbf{Extraction of the signal-to-noise ratio.} \textbf{a}, Signal of the phase output after single demodulation (I) at the tank frequency $f_r$ when the DUT capacitance $C_m=\SI{2}{\femto\farad}$ is continuously connected and disconnected at a \SI{1}{\kilo\hertz} rate.
    \textbf{b}, Fourier power spectrum of $\phi$ with the typical signature of a square wave with exponential transients with the  harmonics $2n+1$.
The signal power spectrum related to the square wave  is isolated in \textbf{d} and the time-domain signal trace is recovered in \textbf{c} by inverted Fourier transform. 
The signal power $P_{sig}$ is computed by integrating the power spectrum of the signal.
    \textbf{f}, Noise power spectrum in $\phi$ extracted as the complementary power spectrum to the signal spectrum shown in \textbf{b}.
    The noise power $P_{noise}$ is computed by integrating the power spectrum of the noise.
    \textbf{e}, Time-domain noise extracted from \textbf{f}.
The signal-to-noise ratio is extracted as the ratio of $P_{sig}/P_{noise}$ and is equal to $5.7$ for an integration time  $t_c$ of \SI{55}{\micro\second} in this example.}
\end{figure}

\begin{figure}[h]
    \includegraphics{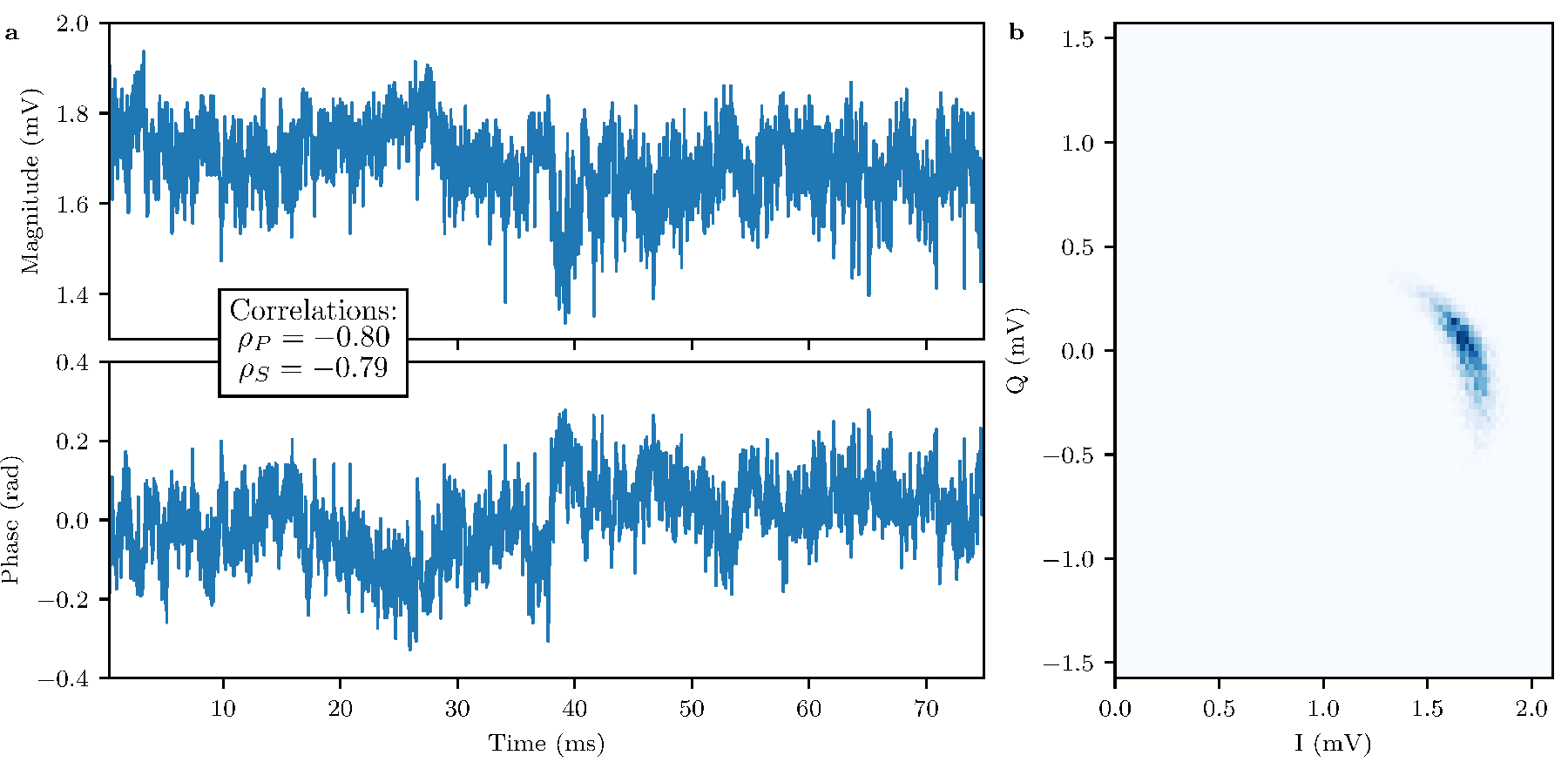}
    \caption{\textbf{Correlated noise in the resonator output signal at 4.2~ K. a}, Magnitude and phase of the output signal after demodulation at the resonance frequency  (method~I).
        Clear correlations between magnitude and phase are observed as a function of time. 
        Both the Spearman $\rho_S$ and Pearson $\rho_P$ computed on the entire time-trace exhibit strong anti-correlation between phase and magnitude with values of about $-0.8$.
        These correlations appear roughly after a  \SI{1}{\milli\second} time-scale. 
    \textbf{b}, 2D histogram in the I-Q plane of the data in \textbf{a}. The non-gaussian banana-shaped spot distribution reflects the evidence for correlated noise in the I-Q plane. This noise is attributed to transistor noise that leads to a noisy inductance value  $L$ and quality factor $Q$.}
\end{figure}

\FloatBarrier
\clearpage
\section{Measurement of the gate capacitance of a different DUT}
\FloatBarrier

\begin{figure}[h]
    \includegraphics{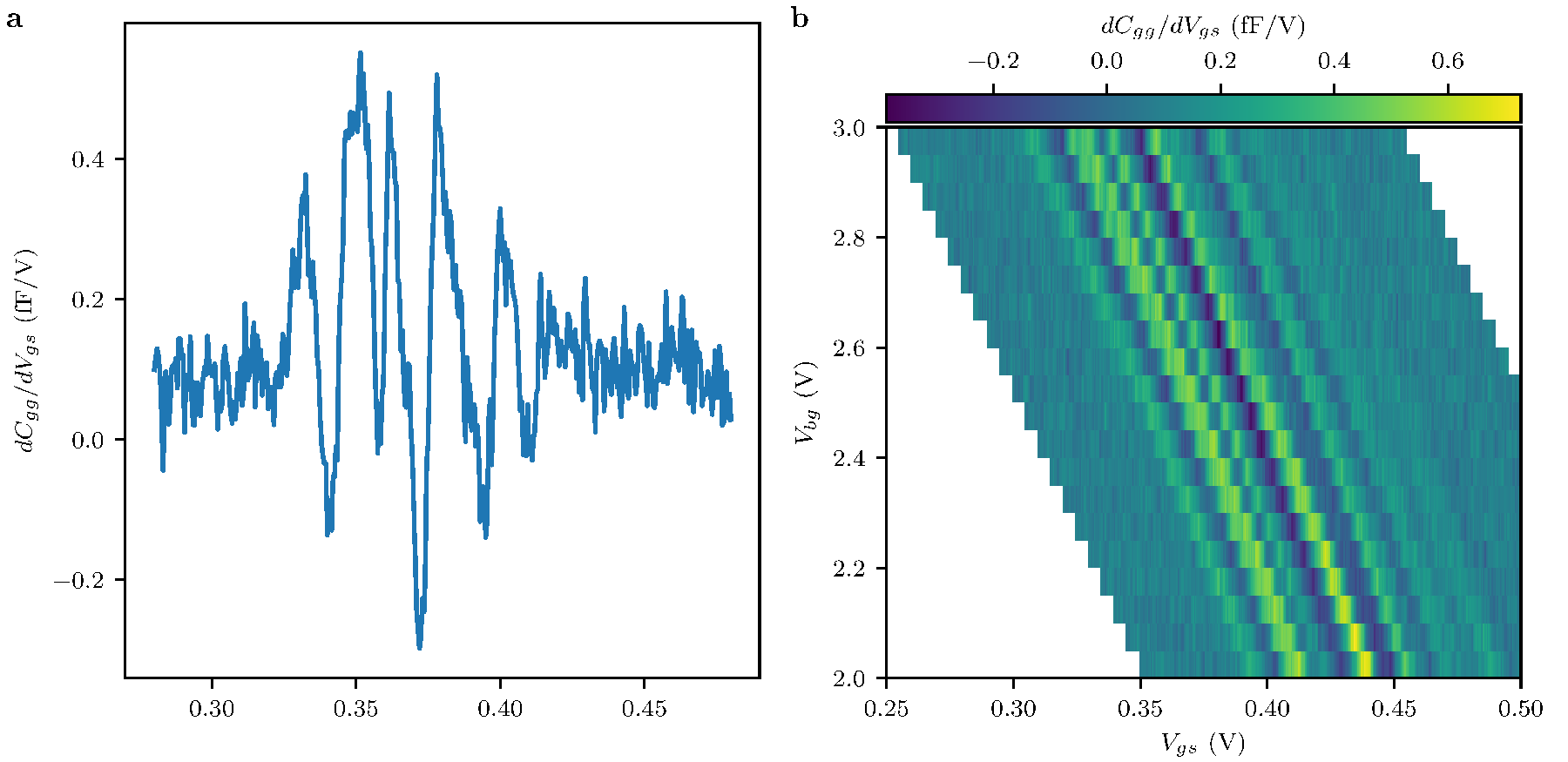}
    \caption{\textbf{Measurement of the N-type DUT transistor with $L=\SI{60}{\nano\meter}$ and $W=\SI{80}{\nano\meter}$. a}, First derivative of the gate capacitance $dC_{gg}/dV_{gs}$ of a N-type MOSFET gate-capacitance as a function of the gate-source voltage $V_{gs}$ measured with the impedancemetry setup at 4.2~K. Oscillations in the derivative capacitance is a sign of quantum capacitance from confined electronic state in the MOSFET channel. 
        \textbf{b}, $dC_{gg}/dV_{gs}$ evolution with front-gate $V_{gs}$ and back-gate $V_{bg}$ voltages. All features shift to lower $V_{gs}$ for increasing $V_{bg}$. 
        No traces attributed to  an impurity state with an anomalous slope  as seen in Figure~5 in the main text are detected for this device. 
    These measurements are the same as presented in the main paper (Figure~5) for a longer device with $L=\SI{120}{\nano\meter}$ and $W=\SI{80}{\nano\meter}$.}
\end{figure}

\FloatBarrier
\clearpage
\section{Experimental setup}
\FloatBarrier
\begin{SCfigure}[10][h]
    \includegraphics[scale=0.8]{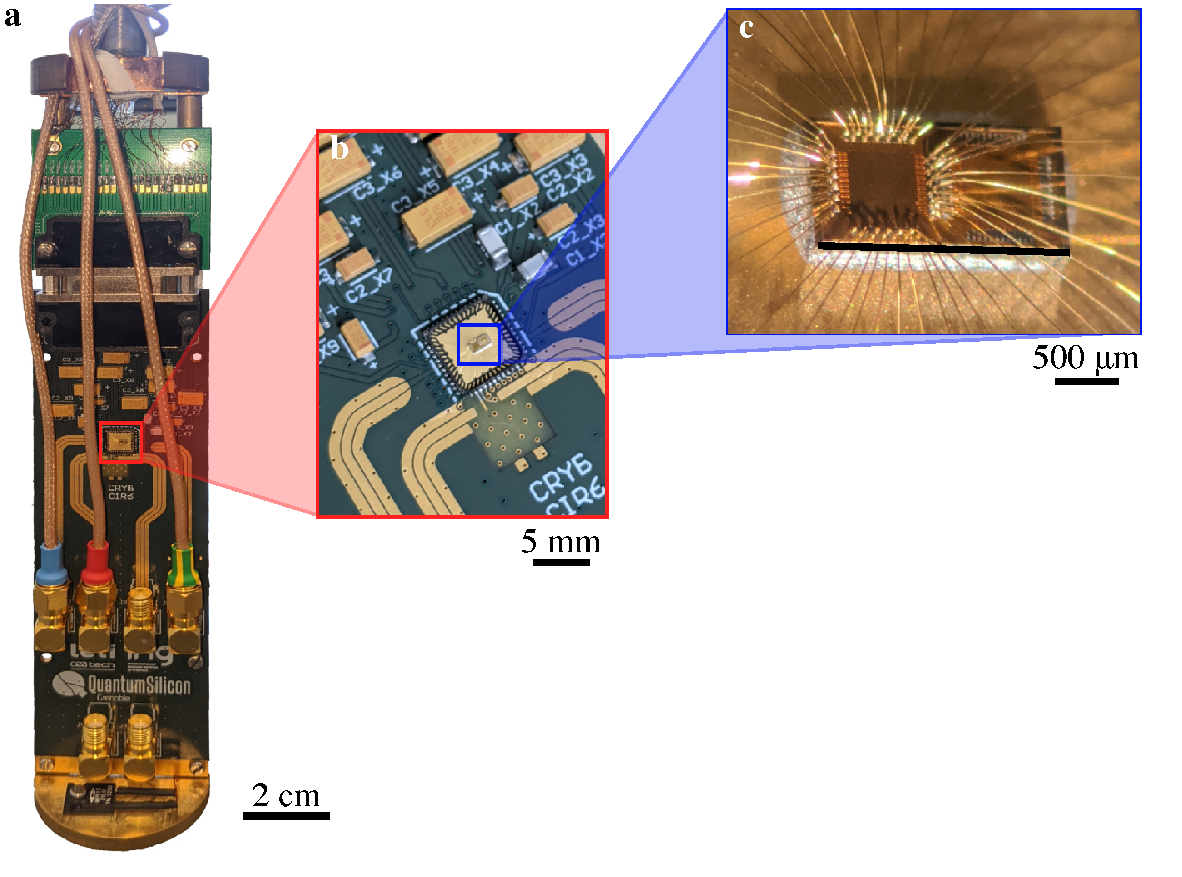}
    \caption{\textbf{Experimental setup of the cryogenic sample holder.} Picture of the PCB (\textbf{a}) with soldered QFN48 (\textbf{b}), and wire-bonded integrated circuit (\textbf{c}).
        The  PCB mounted at the end of a dip-stick is enclosed in a metallic tube filled with He gas for thermal exchange with a liquid helium bath at 4.2~K.
        High frequency signals are routed to SMA connectors at the end of the PCB with grounded coplanar waveguides and conveyed to room temperature with coaxial cables.
        SMD capacitors close to the chip  reject the noise at sensitive voltage nodes ($V_{DD}$, $V_{SS}$,\ldots). 
    }
\end{SCfigure}

\begin{figure}[h]
    \includegraphics[width=0.9\textwidth]{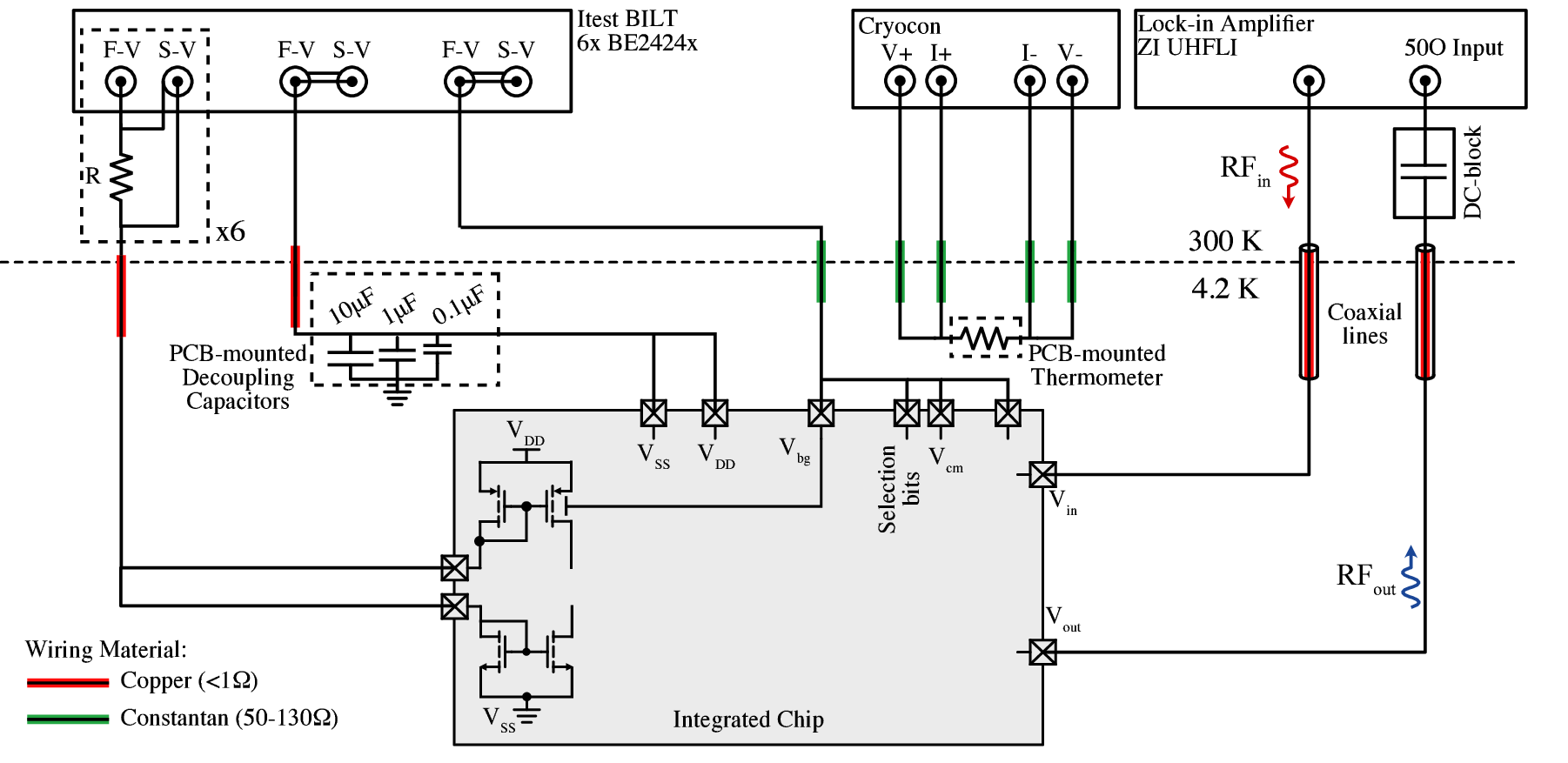}
    \caption{\label{figSM:expsetup}\textbf{Instrumentation with connections to the chip.} The chip is anchored at 4.2~K while all the equipment for voltage and current references,  excitation, and signal detection are placed at room-temperature. 
        The chip power supply lines $V_{DD}$ and $V_{SS}$ are generated from a low-noise voltage source and conveyed with copper wring to avoid voltage drop along the line. 
        Supply voltages are stabilized at cryogenic temperature with 3 discrete SMD capacitors ($2$ ceramic capacitors of 0.1 and \SI{1}{\micro\farad} and 1 tantalum capacitor of \SI{10}{\micro\farad}) placed on the PCB close to the chip pins. 
        Current references are generated by applying a command voltage across a resistor $R$  using the Sense (S-V) and Force (F-V) of the low-noise voltage sources.
        The 6 resistors $R$  of 0.20, 0.40, 0.47, 0.90, 1.4, and \SI{200}{\kilo\ohm} are placed in a grounded metallic enclosure for shielding from environmental noise.
        DC voltages on high-Z inputs (arriving on transistor gates, or back-gate wells) are applied with a low-noise voltage source with constantan wring to reduce heat conduction from 300 to \SI{4.2}{\kelvin}.
        The chip excitation $V_{in}$ and  the output voltage $V_{out}$ conveyed with coaxial SMA lines are generated and treated by a lockin-amplifier with \SI{50}{\ohm}-matched ports.
        A DC block at $V_{out}$ prevents DC currents from the buffer output to flow in the \SI{50}{\ohm} port.
        A thermometer anchored at the PCB ground back-plane is used to monitor the PCB temperature to detect eventual heating above base temperature.
    }
\end{figure}